\documentclass[10pt,prc,preprint,amsmath,amssymb,aps,showpacs,superscriptaddress]{revtex4}
\usepackage{graphicx}
\usepackage{color}

\begin{document}
\preprint{JLAB-THY-18-2753}
\title{ Asymmetric Relativistic Fermi Gas model for quasielastic lepton-nucleus scattering}

\author{M.B. Barbaro}
\affiliation{Dipartimento di Fisica, Universit\`{a} di Torino, Italy}
\affiliation{INFN, Sezione di Torino, Via P. Giuria 1, 10125 Torino, Italy}
\author{A. De Pace}
\affiliation{INFN, Sezione di Torino, Via P. Giuria 1, 10125 Torino, Italy}
\author{T.W. Donnelly}
\affiliation{Center for Theoretical Physics, Laboratory for Nuclear Science and Department of Physics, Massachusetts Institute of Technology, Cambridge, Massachusetts 02139, USA}
\author{J.A. Caballero}
\affiliation{Departamento de F\'{i}sica At\'omica, Molecular y Nuclear, Universidad de Sevilla, 41080 Sevilla, Spain}
\author{G.D. Megias}
\affiliation{Departamento  de F\'{i}sica At\'omica, Molecular y Nuclear, Universidad de Sevilla, 41080 Sevilla, Spain}
\affiliation{IRFU, CEA, Universit\'e Paris-Saclay. F-91191 Gif-sur-Yvette, France}
\author{J. W. Van Orden}
\affiliation{Department of Physics, Old Dominion University, Norfolk, Virginia 23529, USA }
\affiliation{Jefferson Laboratory, 12000 Jefferson Avenue, Newport News, Virginia 23606, USA\footnote{Notice: Authored by Jefferson Science Associates, LLC under U.S. DOE Contract No. DE-AC05-06OR23177.
		The U.S. Government retains a non-exclusive, paid-up, irrevocable, world-wide license to publish or reproduce this manuscript for U.S. Government purposes}}

\begin{abstract}
We develop an asymmetric relativistic Fermi gas model for the study of the electroweak nuclear response in the quasielastic region. The model takes into account the differences between neutron and proton densities in asymmetric $(N>Z)$ nuclei, as well as differences in the neutron and proton separation energies.
We present numerical results for both neutral and charged current processes, focusing on nuclei of interest for ongoing and future neutrino oscillation experiments. We point out some important differences with respect to the commonly employed symmetric Fermi gas model.  
 \end{abstract}


\maketitle

\section{Introduction}
\label{sec:Intro}

We develop the formalism of a fully relativistic Fermi gas (RFG) model for asymmetric nuclear matter and apply this to the study of quasielastic (QE) electron and neutrino scattering on a selected set of nuclei. A similar approach was taken in \cite{Alberico:1989zz,Alberico:1987zgx,Davesne:2014yaa,Lipparini:2013fma,Lipparini:2016fyb}, however using a non-relativistic model, to study various response functions.  Specifically, we focus on a typical $N=Z$ nucleus $^{12}$C, on a slightly asymmetric nucleus $^{40}$Ar, both of which are of practical interest for studies of neutrino oscillations \cite{Alvarez-Ruso:2017oui}, and on a very asymmetric case, $^{208}$Pb. Neutral current electron and (anti)neutrino scattering results are given for these three nuclei, although other cases may be treated in a similar manner. For charge-changing neutrino and antineutrino reactions, of course, the neighboring nuclei $^{12}$B/$^{12}$N, $^{40}$Cl/$^{40}$K, and $^{208}$Tl/$^{208}$Bi, respectively, are also modeled using the same asymmetric Fermi gas approach. Here we focus on QE processes, although the formalism can easily be extended to include the 1p1h inelastic spectrum, namely, meson production, production of baryon resonances and DIS, following the developments in \cite{PhysRevC.69.035502}.  2p2h excitations will be treated in future work.

The paper is organized as follows: in Sect. \ref{sec:Formalism} we introduce the formalism for the asymmetric relativistic Fermi gas (ARFG) model, both for neutral current (Sect. \ref{subsec:NC}) and charged-current (Sect. \ref{subsec:CC}) reactions. In Sect. \ref{sec:Results} we present a selection of numerical results, again for neutral (Sect. \ref{subsec:resNC})  and charged-current (Sect. \ref{subsec:resCC}) reactions. Finally, in Sect. \ref{sec:Concl} we draw our conclusions.

\section{Formalism: the ARFG model}
\label{sec:Formalism}

We consider a nucleus $(N,Z)$ having $N$ neutrons and $Z$ protons, where $N$ and $Z$ need not be equal, and assume that the nuclear volume $V$ for protons and neutrons is the same. Here we are assuming that the cases typically of interest in such studies are in the valley of stability where the equal-volume assumption is a reasonable approximation (in the case of lead, for instance, this is known to be the case); cases far from the valley of stability where, for instance neutron skins may come into play, go beyond the present modeling and are not considered here. Thus we can define two different Fermi spheres for protons and neutrons, with corresponding Fermi momenta
\begin{equation}
k_{F}^{p}(Z) = \left(\frac{3\pi^2 Z}{V}\right)^{1/3} \,,\ \ \ 
k_{F}^{n}(N) = \left(\frac{3\pi^2 N}{V}\right)^{1/3}  \,.
\end{equation}
Defining the ratio between the two Fermi momenta 
\begin{equation}
\rho _{0} \equiv \frac{k_{F}^{n}(N)}{k_{F}^{p}(Z)}=\left( \frac{N}{Z}\right) ^{1/3}  \label{eq1}  
\end{equation}
and the average Fermi momentum
\begin{equation}
  k_{F}^{0}(N,Z) \equiv \frac{1}{N+Z}\left[Z k_{F}^{p}(Z)+N k_{F}^{n}(N)\right] \,,
\label{eq2} 
\end{equation}
we can write
\begin{equation}
k_{F}^{n}(N) = \rho_0 k_{F}^{p}(Z) =  \frac{\rho_0 (Z+N)}{Z+\rho_0 N} k_F^0(N,Z)\,.
\end{equation}
In the study of charged-current (CC) quasielastic neutrino (or antineutrino) scattering, where a neutron is converted into a proton (or {\it vice versa}), one also has to consider 
the neighboring nuclei, (N+1,Z-1) and (N-1, Z+1). The corresponding Fermi momenta for
protons and neutrons can be written as
\begin{eqnarray}
k_{F}^{p}(Z\pm 1) &=& k_{F}^{p}(Z) \left( 1\pm\frac{1}{Z} \right)^{1/3} \\
k_{F}^{n}(N\pm 1) &=& k_{F}^{n}(N) \left( 1\pm\frac{1}{N} \right)^{1/3}\,.
\label{eq5c}
\end{eqnarray}

In the RFG model a nucleon  having 3-momentum $\bf{k}$ has on-shell energy
\begin{equation}
E^{p(n)}(k) = \sqrt{k^{2}+m_{p(n)}^{2}} \,.
\end{equation}
A well-known shortcoming of the RFG consists in the fact that the model has states starting with 3-momentum equal to
zero (energy equal to the proton or neutron mass) and going up to the Fermi
levels. This corresponds to an unrealistic negative separation energy \cite{Cenni:1996zh}. In fact, the Fermi levels in a bound nucleus are negative and have positive separation energies $S_{n}(N)$, $S_{p}(Z)$, $S_{n}(N+1)$, $S_{p}(Z-1)$, $S_{n}(N-1)$ and $S_{p}(Z+1)$, respectively, for the six cases of interest. In order to correct for this flaw, we shift the energies of the protons and neutrons in the triplet of nuclei using the following prescriptions
\begin{equation}
\begin{array}{ll}
H^{n}(N;k)=E^{n}(k)-D^{n}(N) &\ \  H^{p}(Z;k)=E^{p}(k)-D^{p}(Z) \\ 
H^{n}(N+1;k)=E^{n}(k)-D^{n}(N+1)  &\ \  H^{p}(Z-1;k)=E^{p}(k)-D^{p}(Z-1) \\ 
H^{n}(N-1;k)=E^{n}(k)-D^{n}(N-1)  &\ \  H^{p}(Z+1;k)=E^{p}(k)-D^{p}(Z+1)\,,
\end{array}
\label{eq:H}
\end{equation}%
where the offsets are given by%
\begin{equation}
\begin{array}{ll}
D^{n}(N)=E_{F}^{n}(N)+S_{n}(N)  &\ \  D^{p}(Z)=E_{F}^{p}(Z)+S_{p}(Z) \\ 
D^{n}(N+1)=E_{F}^{n}(N+1)+S_{n}(N+1) &\ \  D^{p}(Z-1)=E_{F}^{p}(Z-1)+S_{p}(Z-1) \\ 
D^{n}(N-1)=E_{F}^{n}(N-1)+S_{n}(N-1) &\ \  D^{p}(Z+1)=E_{F}^{p}(Z+1)+S_{p}(Z+1)\,,
\end{array}
\label{eq10b}
\end{equation}%
and where the usual RFG Fermi energies are given by%
\begin{equation}
\begin{array}{ll}
E_{F}^{n}(N)\equiv E^{n}(k_{F}^{n}(N))  &\ \  E_{F}^{p}(Z)\equiv E^{p}(k_{F}^{p}(Z)) \\ 
E_{F}^{n}(N+1)\equiv E^{n}(k_{F}^{n}(N+1))  &\ \  E_{F}^{p}(Z-1)\equiv E^{p}(k_{F}^{p}(Z-1)) \\ 
E_{F}^{n}(N-1)\equiv E^{n}(k_{F}^{n}(N-1))  &\ \  E_{F}^{p}(Z+1)\equiv E^{p}(k_{F}^{p}(Z+1))\,.
\end{array}
\label{eq10c}
\end{equation}%
Clearly when at the true Fermi surfaces the energies in Eq. (\ref{eq:H})
become minus the separation energies. For instance, when $%
k=k_{F}^{n}(N)$ one has%
\begin{equation}
H^{n}(N;k_{F}^{n}(N))=E^{n}(k_{F}^{n}(N))-\left[ E_{F}^{n}(N)+S_{n}(N)\right] =-S_{n}(N).  \label{eq10d}
\end{equation}

In this work the values of the parameter $k_F^0(N,Z)$, in terms of which all the different Fermi momenta can be calculated, are taken from the superscaling analysis \cite{PhysRevC.65.025502} of 
electron scattering data, while $S_{p,n}$ are the measured proton and neutron separation energies, taken from  the ENSDF database \cite{TEPEL1984129}. 
The numerical values for the cases considered in this work are listed in Table I. Note that, although not explicitly indicated in our notation for sake of simplicity, all separation energies depend on both $Z$ and $N$.
\begin{table}[h]
  \begin{tabular}{| c | c | c | c | c | c | c | c |}
    \hline
    X(A,Z,N)     &  $S_n$ (MeV)  & $S_p$ (MeV)  &  $k_F^0$ (MeV/c)  &  $k_F^n$ (MeV/c)  &  $k_F^p$ (MeV/c)   \\ \hline \hline
 C(12,6,6)       &      18.72    &        15.96  & 228 &      228 & 228    \\ \hline
 B(12,5,7)       &       3.37    &        14.10  &   & 240.02 & 214.56  \\ \hline
 N(12,7,5)       &      15.04    &         0.60  &  & 214.56 & 240.02  \\ \hline \hline
Ar(40,18,22)     &       9.87    &        12.53  & 241 &    248.23 & 232.17  \\ \hline
Cl(40,17,23)     &       5.83    &        11.68  &  & 251.93 & 227.78 \\ \hline
 K(40,19,21)     &       7.80    &         7.58  &  & 244.41 & 236.39  \\ \hline \hline
Pb(208,82,126)   &       7.37    &         8.00  & 248 &  261.77 & 226.85    \\ \hline
Tl(208,81,127)   &       3.79    &         7.55  &  & 262.46 & 225.92 \\ \hline
Bi(208,83,125)   &       6.89    &         3.71  &  & 261.07 & 227.76 \\ \hline
  \end{tabular}
  \label{tab:tab1}
\caption{Neutron and proton separation energies ($S_n$, $S_p$) and Fermi momenta ($k_F$) used in this work. 
}
\end{table}

Within this model, denoted as Asymmetric Relativistic Fermi Gas (ARFG), we can now calculate the quasielastic double differential cross section with respect to the outgoing lepton momentum $k'$ and scattering angle $\Omega$ corresponding to inclusive electron scattering, $(e,e')$, neutral current (NC) neutrino and antineutrino scattering, $(\nu,\nu')$ and $(\bar\nu,\bar\nu')$, and to charged-current (CC) neutrino and antineutrino scattering, $(\nu,\mu^-)$ and $(\bar\nu,\mu^+)$. 
In the Rosenbluth decomposition these can be expressed as
\begin{eqnarray}
\left(\frac{d^2\sigma}{d\Omega dk'}\right)^{(e,e')} &=& \sigma_M \left[v_L R^{L,em}_{p\to p}+  v_L R^{L,em}_{n\to n} +v_T R^{T,em}_{p\to p} +
v_T R^{T,em}_{n\to n}\right] 
  \label{eq:eep}\\
\left(\frac{d^2\sigma}{d\Omega dk'}\right)^{(\nu,\nu')} &=& \sigma_0^{(NC)} \left[v_{L} R^{L,w}_{n\to n} +  v_T R^{T,w}_{n\to n} + v_{T'} R^{T',w}_{n\to n} \right]
  \label{eq:nunu}\\
\left(\frac{d^2\sigma}{d\Omega dk'}\right)^{(\bar\nu,\bar\nu')} &=& \sigma_0^{(NC)} \left[v_{L} R^{L,w}_{p\to p} +  v_T R^{T,w}_{p\to p} - v_{T'} R^{T',w}_{p\to p} \right]
  \label{eq:nubarnubar}\\
\left(\frac{d^2\sigma}{d\Omega dk'}\right)^{(\nu,\mu^-)} &=& \sigma_0^{(CC)} \left[V_{CC} R^{CC,w}_{n\to p} +V_{CL} R^{CL,w}_{n\to p} +V_{LL} R^{LL,w}_{n\to p} +
  V_T R^{T,w}_{n\to p} + V_{T'} R^{T',w}_{n\to p}\right]
    \label{eq:numu}\\
\left(\frac{d^2\sigma}{d\Omega dk'}\right)^{(\bar\nu,\mu^+)} &=& \sigma_0^{(CC)} \left[V_{CC} R^{CC,w}_{p\to n} +V_{CL} R^{CL,w}_{p\to n}+V_{LL} R^{LL,w}_{n\to n}+
  V_T R^{T,w}_{p\to n}- V_{T'} R^{T',w}_{p\to n} \right]\,,
  \label{eq:nubarmu}
\end{eqnarray}
where $v_K$ and $V_K$ are leptonic kinematic factors (see \cite{Amaro:2004bs,PhysRevC.73.035503} for their explicit expressions), $\sigma_M$ is the Mott cross section and 
$\sigma_0^{(NC)}$, $\sigma_0^{(CC)}$ the corresponding elementary weak cross sections for NC and CC reactions, respectively.
The response functions $R^K\equiv R^K(q,\omega)$, where the labels $em$ and $w$ stand for ``electromagnetic" and ``weak", respectively, embody the nuclear structure and dynamics and are functions of the momentum $q$ and energy $\omega$ transferred to the nucleus. They are related to the specific components of the corresponding hadronic tensor $W^{\mu\nu}$:
\begin{eqnarray}
&& R_L \equiv R_{CC} = W^{00}
\\
&& R_{CL} = - \frac{1}{2} \left( W^{03} + W^{30} \right)
\\
&& R_{LL} = W^{33} 
\\
&& R_{T} = W^{11} + W^{22} 
\\
&& R_{T'} = - \frac{i}{2} \left( W^{12} - W^{21} \right) \,.
\end{eqnarray}
The general expression for the nuclear tensor in the ARFG model is
\begin{equation}
W^{\mu\nu}_{i\to f}(q,\omega)
=
\frac{3 m^2  {\cal N}}{4\pi (k_F^i)^3}
\int d{\bf h} \,
\frac{\theta(k_F^i-|{\bf h}|)\,\theta(|{\bf h}+{\bf q}|-k_F^f)
}{H^i({\bf h}) H^f({\bf h}+{\bf q})} f^{\mu\nu}_{i\to f}({\bf h},{\bf h}+{\bf q})
\, \delta[H^f({\bf h}+{\bf q})-H^i({\bf h})-\omega] \,,
\label{eq:Wmunu}
\end{equation}
where the superscripts $i$ and $f$ refer to the initial and final nucleons, respectively, $m$ and ${\cal N}$ are the appropriate mass and number of nucleons in the target nucleus, $H^{i,f}$ are the nucleon energies defined in Eqs. (\ref{eq:H}) and $f^{\mu\nu}_{i\to f}$ is the corresponding single-nucleon tensor.

In the following subsections we shall derive the explicit expression of $W^{\mu\nu}$ for the reactions listed in Eqs. (\ref{eq:eep}-\ref{eq:nubarmu}), distinguishing between the two cases of neutral and charged-current reactions.

\subsection{Electron Scattering and NC (Anti)Neutrino Scattering}
\label{subsec:NC}

In the case of neutral current processes, Eqs. (\ref{eq:eep}-\ref{eq:nubarnubar}), mediated by the exchange of a photon or a $Z^0$ boson, the energy-conserving delta-function appearing in Eq.~(\ref{eq:Wmunu})  involves the difference between the {\it on-shell}
particle (p) and hole (h) energies, where the hole is assumed to have
3-momentum $\bf{h}$, while the particle has 3-momentum $\bf{h}+\bf{q}$,
\begin{eqnarray}
E^{n\rightarrow n}(N) &=&H^{n}(N;{\bf h}+{\bf q})-H^{n}(N;h) = E^{n}({\bf h}+{\bf q})-E^{n}(h)  \label{eq12} \\
E^{p\rightarrow p}(Z) &=&H^{p}(Z;{\bf h}+{\bf q})-H^{p}(Z;h) = E^{p}({\bf h}+{\bf q})-E^{p}(h)\,,  \label{eq14}
\end{eqnarray}%
since the $D^{n}(N)$ and $D^{p}(N)$ offsets cancel in the particle-hole
energy differences; namely, the nucleon separation energies introduced in the ARFG model have no impact on the results for neutral current processes. The only difference with respect to the usual RFG arises from the different values of $k_F$ for protons and neutrons.
The corresponding nuclear tensors are then
\begin{equation}
W^{\mu\nu}_{p\rightarrow p} (q,\omega)
=
\frac{3 m_p^2  Z}{4\pi \left[k_{F}^{p}(Z)\right]^3}
\int d{\bf h} \,
\frac{\theta(k_{F}^{p}(Z)-|{\bf h}|)\,\theta(|{\bf h}+{\bf q}|-k_{F}^{p}(Z))
}{E^{p}({\bf h}) E^{p}({\bf h}+{\bf q})}
f^{\mu\nu}_{p\rightarrow p}({\bf h},{\bf h}+{\bf q})
\, \delta[E^{p}({\bf h}+{\bf q})-E^{p}({\bf h})-\omega]
\label{eq:Wmunup}
\end{equation}
for protons and
\begin{equation}
W^{\mu\nu}_{n\rightarrow n} (q,\omega)
=
\frac{3 m_n^2  N}{4\pi \left[k_{F}^{n}(N)\right]^3}
\int d{\bf h} \,
\frac{\theta(k_{F}^{n}(N)-|{\bf h}|)\,\theta(|{\bf h}+{\bf q}|-k_{F}^{n}(N))
}{E^{n}({\bf h}) E^{n}({\bf h}+{\bf q})}
f^{\mu\nu}_{n\rightarrow n}({\bf h},{\bf h}+{\bf q})
\, \delta[E^{n}({\bf h}+{\bf q})-E^{n}({\bf h})-\omega]
\label{eq:Wmunun}
\end{equation}
for neutrons, where
\begin{eqnarray}
f^{\mu\nu}_{j\rightarrow j} &=& - w_{1,j}(\tau_j) \left( g^{\mu\nu} - \frac{Q^\mu Q^\nu}{Q^2} \right)
+ w_{2,j}(\tau_j) V^\mu_j V^\nu_j - \frac{i}{m_n} w_{3,n}(\tau_n) \epsilon^{\mu\nu\rho\sigma}
Q_\rho V_{\sigma,j}
\label{eq:fmunuj}
\end{eqnarray}
(with $j=p,n$) are the single-nucleon tensors, with
\begin{equation}
  \tau_j \equiv \frac{|Q^2|}{4m_j^2}
  \ \ \ \mbox{and} \ \ \   V^\mu_j = \frac{1}{m_j} \left( P^\mu - \frac{P\cdot Q}{Q^2}\, Q^\mu \right)
  = \frac{1}{m_j} \left( P^\mu + \frac{1}{2}\, Q^\mu \right)\,,
\end{equation}
having used the on-shell condition $\frac{P\cdot Q}{Q^2}=-\frac{1}{2}$.

For $(e,e')$ the electromagnetic structure functions are
\begin{eqnarray}
  w_{1,j}(\tau) &=& \tau G_{M,j}^2(\tau)
  \\
  w_{2,j}(\tau) &=& \frac{G_{E,j}^2(\tau) + \tau G_{M,j}^2(\tau)}{1+\tau}
  \label{eq:wem}
\end{eqnarray}
and $w_3=0$, while for $(\nu,\nu')$ and $(\bar\nu,\bar\nu')$ they are 
\begin{eqnarray}
  w_{1,j}(\tau) &=& \tau \widetilde G_{M,j}^2(\tau) + (1+\tau) \widetilde G_{A,j}^2(\tau)
  \\
  w_{2,j}(\tau) &=& \frac{\widetilde G_{E,j}^2(\tau) + \tau \widetilde G_{M,j}^2(\tau)}{1+\tau} + \widetilde G_{A,j}^2(\tau)
  \\
  w_{3,j}(\tau) &=& \widetilde G_{M,j}(\tau)\widetilde G_{A,j}(\tau) \,,
\label{eq:wnc}
\end{eqnarray}
with $\tau=\tau_{n,p}$ as is appropriate.
By performing the angular integration in Eqs.~(\ref{eq:Wmunup}) and (\ref{eq:Wmunun})  one gets 
\begin{eqnarray}
W^{\mu\nu}_{p\rightarrow p} (q,\omega)
&=&
\frac{3 m_p^2 Z}{2 \left[k_{F}^{p}(Z)\right]^3 q}
\int_{E_{0}^{p}(Z)}^{E_{F}^{p}(Z)} dE 
\left[f^{\mu\nu}_{p\rightarrow p}\right]_{x=x_0^p(E)} \,
\\
W^{\mu\nu}_{n\rightarrow n} (q,\omega)
&=&
\frac{3 m_n^2 N}{2 \left[k_{F}^{n}(N)\right]^3 q}
\int_{E_{0}^{n}(N)}^{E_{F}^{n}(N)} dE 
\left[f^{\mu\nu}_{n\rightarrow n}\right]_{x=x_0^n(E)} \,,
\end{eqnarray}
where
\begin{eqnarray}
 E_{0}^{p}(Z) &=& \max\left\{E_{F}^{p}(Z)-\omega ,
\frac{q}{2} \sqrt{1+\frac{1}{\tau_p}}-\frac{\omega}{2}
\right\} 
\label{eq:eps0p}
\\
 E_{0}^{n}(N) &=& \max\left\{E_{F}^{n}(N)-\omega ,
  \frac{q}{2} \sqrt{1+\frac{1}{\tau_n}}-\frac{\omega}{2}
\right\} 
\label{eq:eps0n}
\end{eqnarray}
and
\begin{equation}
  x_0^{n,p}(E)   = \frac{\omega E-\frac{|Q^2|}{2}}{q \sqrt{E^2-m_{n,p}^2}} \,.
\end{equation}

Finally, the analytic integration over $E$ yields:
\begin{eqnarray}
  W^{\mu\nu}_{p\to p}(q,\omega) &=& \frac{3 m_p^2 Z}{2 \left[k_{F} (Z) \right]^3 q} \, (E_F^p(Z)-E_0^p(Z))\,U_p^{\mu\nu}
  \\
  W^{\mu\nu}_{n\to n}(q,\omega) &=& \frac{3 m_n^2 N}{2 \left[k_{F} (N) \right]^3 q} \, (E_F^n(N)-E_0^n(N))\,U_n^{\mu\nu} \,.
\end{eqnarray}
In particular, the relevant components for the calculation of the L, T and T' responses turn out to be
\begin{eqnarray}
  U^{00}_j &=& \frac{\kappa^2}{\tau_j} \left[ -w_{1,j}(\tau_j) + (1+\tau_j) w_{2,j}(\tau_j)
    +  w_{2,j} (\tau_j) \Delta_j \right]
  \\
  U^{11}_j + U^{22}_j &=& 2  w_{1,j}(\tau_j) +   w_{2,j} (\tau_j) \Delta_j
  \\
  U^{12}_j &=& 2 i \sqrt{\tau_j(1+\tau_j)}\,  w_{3,j}(\tau_j) (1 + \Delta_j^\prime)
\end{eqnarray}
with
\begin{eqnarray}
  \Delta_j &=& \frac{\tau_p}{\kappa^2} \frac{1}{m_j^2} \left[\frac{1}{3} \left(E_F^{j2}(Z)+E_0^j(Z) E_F^j(Z)
    +E_0^{j2}(Z)\right) + \frac{\omega}{2} \left(E_F^j(Z)+E_0^j(Z)\right) + \frac{\omega^2}{4}\right] -1 -\tau_j \\
  \Delta^\prime_j &=& \frac{1}{\kappa} \sqrt{\frac{\tau_j}{1+\tau_j}} \frac{1}{m_j} \left[
    \frac{\omega}{2} + \frac{1}{2} \left(E_F^j(Z)+E_0^j(Z)\right)\right] - 1 \,.
  \end{eqnarray}
The corresponding  numerical results will be shown in Sect. \ref{sec:Results}.

\subsection{CC (Anti)Neutrino Scattering}
\label{subsec:CC}

In the case of (anti)neutrino-induced CC reactions, where neutrons (protons) are converted
into protons (neutrons) through the absorption of a $W^+$ ($W^-$) boson, the energy differences appearing in the delta-function of Eq. (\ref{eq:Wmunu}) are 
\begin{eqnarray}
E^{n\rightarrow p}(N,Z;ph) &=&H^{p}(Z+1;p)-H^{n}(N;h)
\nonumber \\
&=&\left[ E^{p}(p)-E^{n}(h)\right] +\Delta D^{n\rightarrow p}(N,Z),
\label{eq16}
\end{eqnarray}%
with%
\begin{eqnarray}
\Delta D^{n\rightarrow p}(N,Z) &\equiv &D^{n}(N)-D^{p}(Z+1)  \nonumber \\
&=&\left[ E_{F}^{n}(N)-E_{F}^{p}(Z+1)\right] +\left[S_n(N)-S_{p}(Z+1)\right]   \label{eq17a}
\end{eqnarray}%
for CC neutrino reactions, and%
\begin{eqnarray}
E^{p\rightarrow n}(N,Z;k^{\prime }k) &=&H^{n}(N+1;k^{\prime })-H^{p}(Z;k)
\nonumber \\
&=&\left[ E^{n}(k^{\prime })-E^{p}(k)\right] +\Delta D^{p\rightarrow n}(N,Z) \,,
\label{eq19}
\end{eqnarray}%
with%
\begin{eqnarray}
\Delta D^{p\rightarrow n}(N,Z) &\equiv &D^{p}(Z)-D^{n}(N+1)  \nonumber \\
&=&\left[ E_{F}^{p}(Z)-E_{F}^{n}(N+1)\right] +\left[ \varepsilon
_{s}^{p}(Z)-\varepsilon _{s}^{n}(N+1)\right]   \label{eq20a}
\end{eqnarray}%
for CC antineutrino reactions. The numerical values for these energy offsets are given in Table II.

\begin{table}[h]
  \begin{tabular}{| c | c | c |}
    \hline
    X(A,Z,N)     &   $D^{n\to p}$ (MeV)  &  $D^{p\to n}$ (MeV) \\ \hline \hline
 C(12,6,6)     &    15.21  & 9.68    \\ \hline
Ar(40,18,22)    &  5.25   &   1.77   \\ \hline
Pb(208,82,126) & 12.24 &  -4.77   \\ \hline
  \end{tabular}
\label{tab:tab2}
\caption{Energy offsets  used in this work for CC neutrino ($D^{n\to p}$) and antineutrino ($D^{p\to n}$) scattering. }
\end{table}

The corresponding nuclear tensors are then
\begin{eqnarray}
W^{\mu\nu}_{n\rightarrow p} (q,\omega)
&=&
\frac{3 m_n^2  N}{4\pi \left[k_{F}^{n}(N)\right]^3}
\int d{\bf h} \,
\frac{\theta(k_{F}^{n}(N)-|{\bf h}|)\,\theta(|{\bf h}+{\bf q}|-k_{F}^{p}(Z+1))
}{E^{n}({\bf h}) E^{p}({\bf h}+{\bf q})}
\nonumber\\ &\times&
f^{\mu\nu}_{n\rightarrow p}({\bf h},{\bf h}+{\bf q})
\, \delta[E^{p}({\bf h}+{\bf q})-E^{n}({\bf h})+\Delta D^{n\rightarrow p}(N,Z) -\omega]
\\
W^{\mu\nu}_{p\rightarrow n} (q,\omega)
&=&
\frac{3 m_p^2  Z}{4\pi \left[k_{F}^{p}(Z)\right]^3}
\int d{\bf h} \,
\frac{\theta(k_{F}^{p}(Z)-|{\bf h}|)\,\theta(|{\bf h}+{\bf q}|-k_{F}^{n}(N-1))
}{E^{p}({\bf h}) E^{n}({\bf h}+{\bf q})}
\nonumber\\ &\times&
 f^{\mu\nu}_{p\rightarrow n}({\bf h},{\bf h}+{\bf q})
\, \delta[E^{n}({\bf h}+{\bf q})-E^{p}({\bf h})+\Delta D^{p\rightarrow n}(N,Z)-\omega]
\nonumber\\
\end{eqnarray}
for neutrino and antineutrino scattering, respectively, where the elementary isovector tensor $f^{\mu\nu}_{n\rightarrow p} = f^{\mu\nu}_{p\rightarrow n} \equiv f^{\mu\nu (1)} $ is
\begin{equation}
  f^{\mu\nu (1)} =
  - w_{1}^{(1)}(\tau) \left( g^{\mu\nu} - \frac{Q^\mu Q^\nu}{Q^2} \right)
  + w_{2}^{(1)}(\tau) V^\mu V^\nu + u_{1}^{(1)}(\tau) \frac{Q^\mu Q^\nu}{Q^2}
  - \frac{i}{m} w_{3}^{(1)}(\tau) \epsilon^{\mu\nu\rho\sigma} Q_\rho V_{\sigma}
\label{eq:fmunu1}
\end{equation}
and the structure functions $w_i$ are the appropriate isovector ones:
\begin{eqnarray}
  w_{1}^{(1)}(\tau) &=& \tau \left[ G_{M}^{(1)} (\tau) \right]^2
  + (1+\tau) \left[ G_{A}^{(1)} (\tau) \right]^2
  \\
  w_{2}^{(1)}(\tau) &=& \frac{\left[ G_{E}^{(1)}(\tau)\right]^2 + \tau \left[G_{M,i}^{(1)}(\tau)\right]^2}{1+\tau} + \left[ G_{A}^{(1)} (\tau) \right]^2
  \\
  u_{1}^{(1)}(\tau) &=& - \left[ G_{A}^{\prime (1)}(\tau) \right]^2
  \\
  w_{3}^{(1)}(\tau) &=& G_{M}^{(1)}(\tau) G_{A}^{(1)}(\tau) \,,
\label{eq:wcc}
\end{eqnarray}
where
\begin{equation}
  G_{A}^{\prime (1)}(\tau) = G_{A}^{(1)}(\tau) - \tau G_{P}^{(1)}(\tau) \,.
  \end{equation}
In the case of CC$\nu$ reactions we set
$m_n\cong m_p\cong m\equiv (m_n + m_p)/2$ and define a single dimensionless 4-momentum transfer $\tau\equiv |Q^2|/4m^2$.

Similarly to what we did for the NC case, we perform the angular integral,  obtaining
\begin{equation}
W^{\mu\nu}_{n\rightarrow p} (q,\omega)
=
\frac{3 m_n^2 N}{2 \left[k_{F}^{n}(N)\right]^3 q}
\int_{E_{0}^{n\rightarrow p}(N)}^{E_{F}^{n}(N)} dE 
\left[f^{\mu\nu}_{n\rightarrow p}\right]_{x=x_0^{n\rightarrow p}(E)} \,
\label{eq:Wmununp}
\end{equation}
\begin{equation}
W^{\mu\nu}_{p\rightarrow n} (q,\omega)
=
\frac{3 m_p^2 Z}{2 \left[k_{F}^{p}(Z)\right]^3 q}
\int_{E_{0}^{p\rightarrow n}(Z)}^{E_{F}^{p}(Z)} dE 
\left[f^{\mu\nu}_{p\rightarrow n}\right]_{x=x_0^{p\rightarrow n}(E)} \,,
\label{eq:Wmunupn}
\end{equation}
where now
\begin{equation}
  x_0^{n\rightarrow p}(E)   = \frac{\widetilde\omega_{n\rightarrow p} E-\frac{|\widetilde Q^2_{n\rightarrow p}|}{2}}{q \sqrt{E^2-m_{n}^2}} 
  \ \ \ \mbox{and}\ \ \
     x_0^{p\rightarrow n}(E)   = \frac{\widetilde\omega_{p\rightarrow n} E-\frac{|\widetilde Q^2_{p\rightarrow n}|}{2}}{q \sqrt{E^2-m_{p}^2}} \,,
\label{eq:x0nppn}
\end{equation}
having defined
\begin{eqnarray}
  \widetilde\omega_{n\rightarrow p} &\equiv&
  \omega-\Delta D^{n\rightarrow p}(N,Z)
  \ \ \ \mbox{and}\ \ \
  \widetilde Q^2_{n\rightarrow p} \equiv \widetilde\omega_{n\rightarrow p}^2-q^2 
  \label{eq:omtil-np}
  \\
  \widetilde\omega_{p\rightarrow n} &\equiv&
  \omega-\Delta D^{p\rightarrow n}(N,Z)
  \ \ \ \mbox{and}\ \ \
  \widetilde Q^2_{p\rightarrow n} \equiv \widetilde\omega_{p\rightarrow n}^2-q^2 \,.
  \label{eq:omtil-pn}
\end{eqnarray}

The lower limits of integration in Eqs. (\ref{eq:Wmununp},\ref{eq:Wmunupn}) are
\begin{eqnarray}
  E_{0}^{n\rightarrow p}(N) &=& \max\left\{  E_{F}^{p}(Z+1)-\widetilde\omega_{n\rightarrow p} ,
  \Gamma^{n\rightarrow p}   \right\} 
\label{eq:eps0ntop}
\\
  E_{0}^{p\rightarrow n}(Z) &=& \max\left\{ E_{F}^{n}(N+1)-\widetilde\omega_{p\rightarrow n} ,
  \Gamma^{p\rightarrow n}   \right\} 
\label{eq:eps0pton}
\end{eqnarray}
with
\begin{equation}
  \Gamma^{n\rightarrow p} =  \frac{q}{2} \sqrt{1 + \frac{1}{\widetilde\tau_{n\rightarrow p}}}
  -\frac{\widetilde\omega_{n\rightarrow p}}{2} 
  \ \ \ \mbox{and}\ \ \
  \Gamma^{p\rightarrow n} =  \frac{q}{2} \sqrt{1 + \frac{1}{\widetilde\tau_{p\rightarrow n}}}
  -\frac{\widetilde\omega_{p\rightarrow n}}{2}\,,
\end{equation}
together with
${\widetilde\tau}_{n\rightarrow p}\equiv \frac{\left|\widetilde Q^2_{n\rightarrow p}\right|}{4m^2}$
and
${\widetilde\tau}_{p\rightarrow n}\equiv \frac{\left|\widetilde Q^2_{p\rightarrow n}\right|}{4m^2}$. 

Although in principle also in this case, as in the NC one, it is possible to obtain fully analytic results,  
for practical purposes it is easier to perform the energy integral numerically. The corresponding results will be shown in the next section.

\section{Results}
\label{sec:Results}

In this section we present and compare the nuclear response functions evaluated in the symmetric (SRFG) and asymmetric (ARFG) relativistic Fermi gas models.

As anticipated, in the case of neutral current reactions, $(e,e')$, $(\nu,\nu')$ and $(\bar\nu,\bar\nu')$,  the ARFG results differ from the SRFG ones only due to the different neutron and proton Fermi momenta. On the contrary, for charged-current reactions,  $(\nu,\mu^-)$ and $(\bar\nu,\mu^+)$, the energy offsets related to the different separation energies in the initial and final nuclei also play a role.

\subsection{Neutral current reactions}
\label{subsec:resNC}

\begin{figure}[h]
\includegraphics[scale=0.8, angle=0]{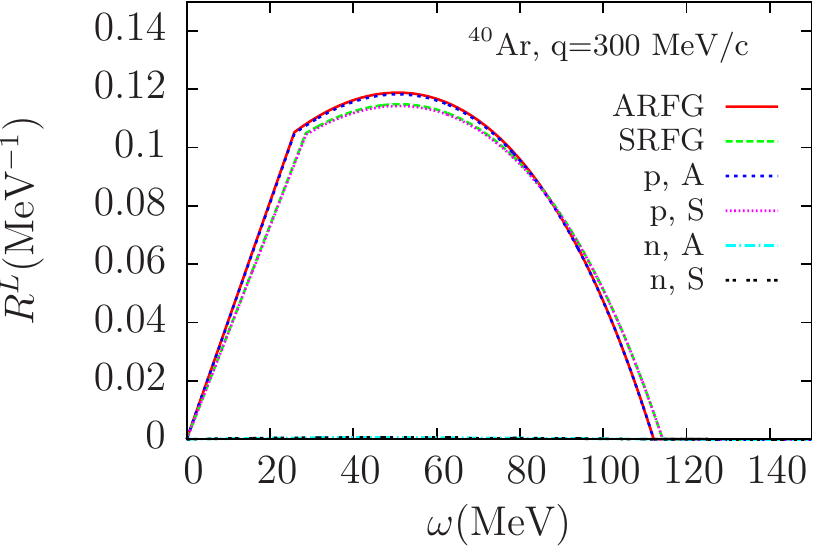}
\hspace{1.cm}
\includegraphics[scale=0.8, angle=0]{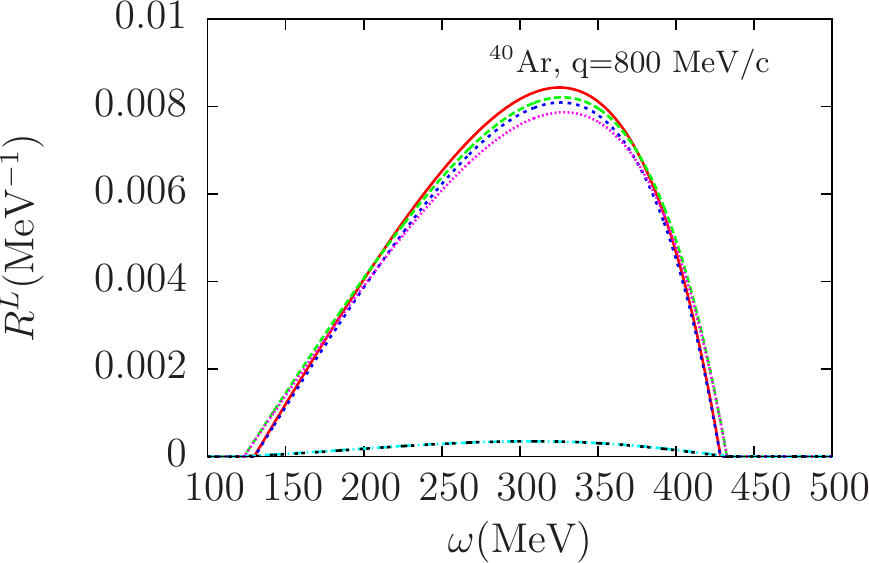}
\\
\includegraphics[scale=0.8, angle=0]{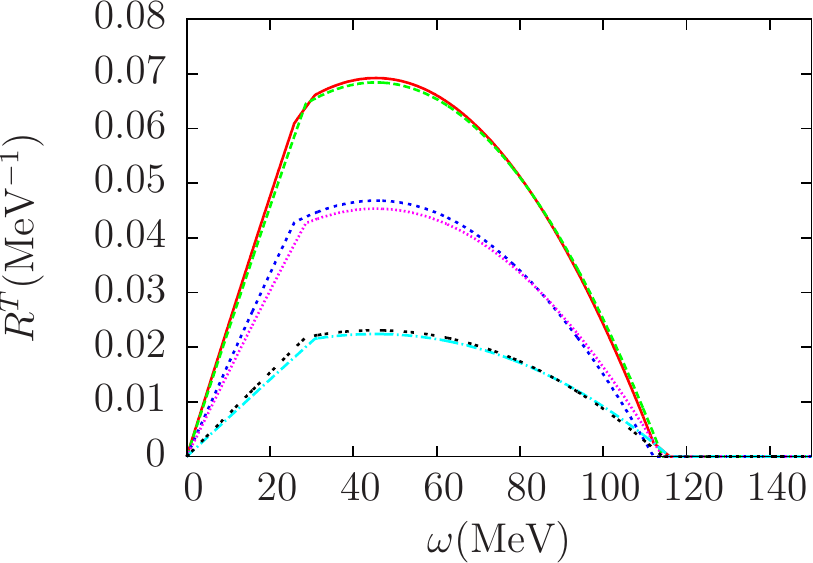}
\hspace{1.cm}
\includegraphics[scale=0.8, angle=0]{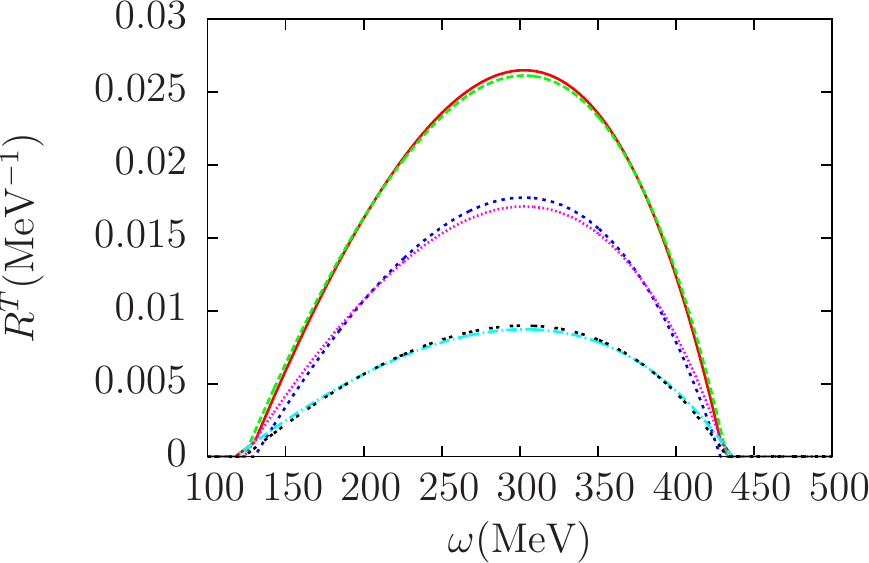}
\caption{(Color online) Electromagnetic response functions of $^{40}$Ar 
 in the symmetric (SRFG) and asymmetric (ARFG) relativistic Fermi gas for momentum transfer $q=300$ (left column) and 800 (right column) MeV/c.
The separate contributions of protons and neutrons are also displayed.}
\label{fig:Aree} 
\end{figure}
\begin{figure}[h]
\includegraphics[scale=0.8, angle=0]{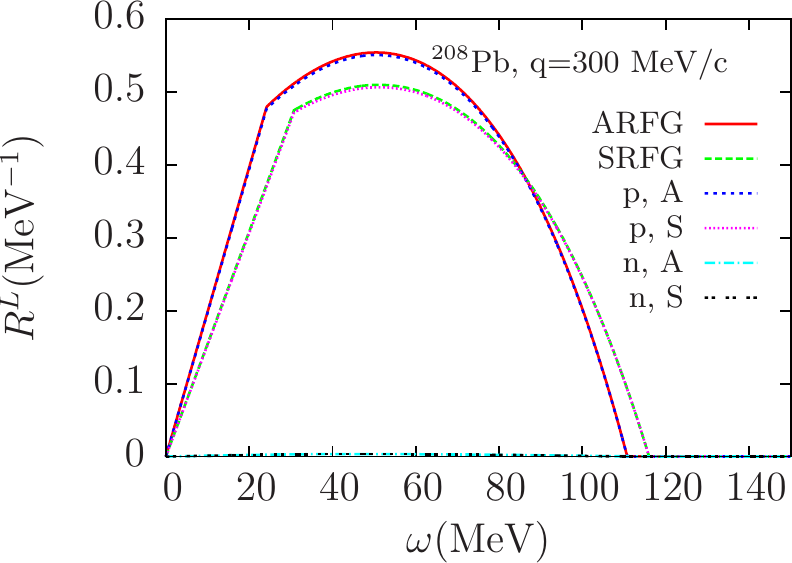}
\hspace{1.cm}
\includegraphics[scale=0.8, angle=0]{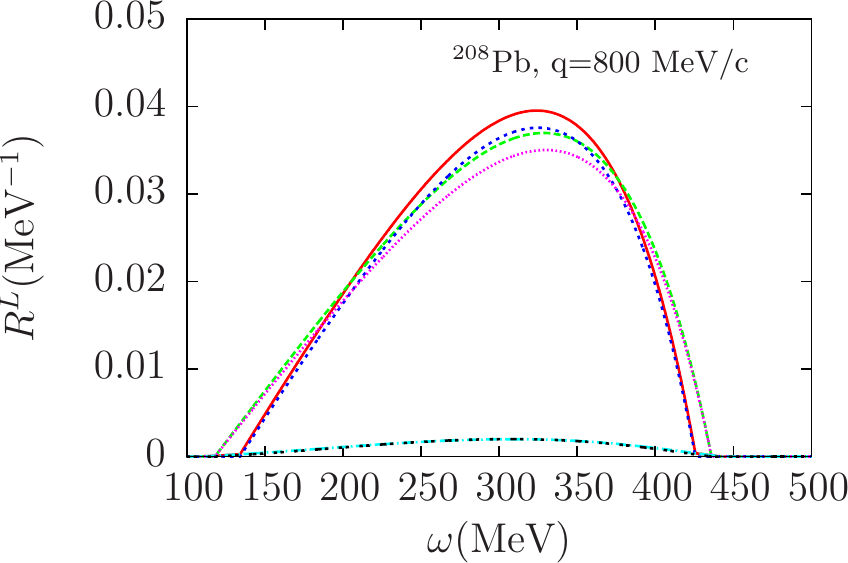}
\\
\includegraphics[scale=0.8, angle=0]{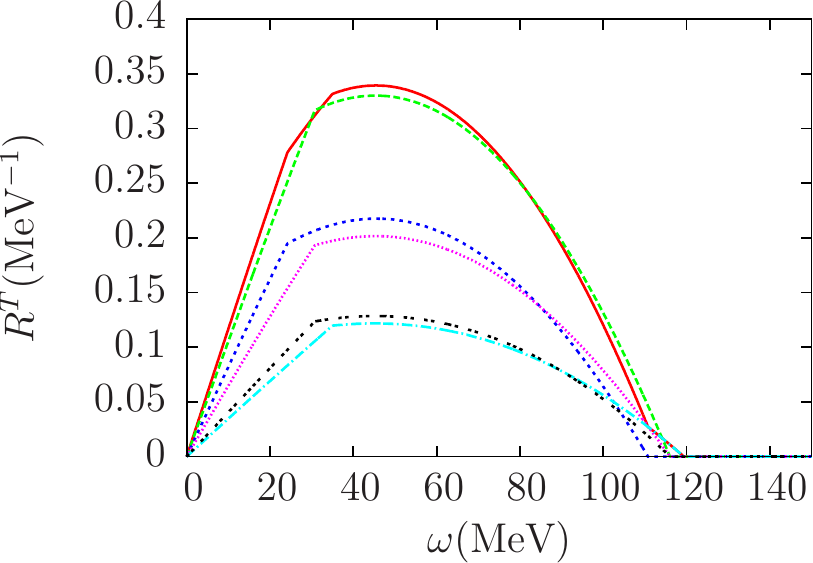}
\hspace{1.cm}
\includegraphics[scale=0.8, angle=0]{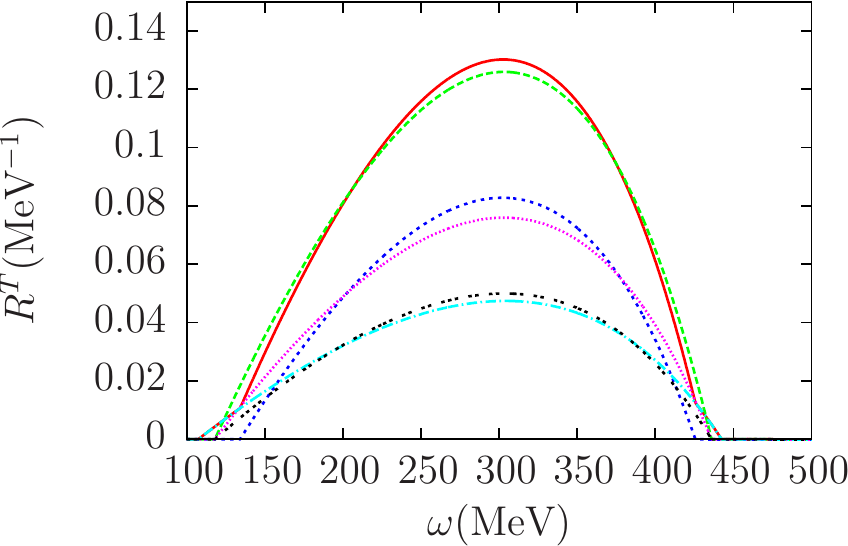}
\caption{(Color online) Electromagnetic response functions of $^{208}$Pb 
 in the symmetric (SRFG) and asymmetric (ARFG) relativistic Fermi gas.
The separate contributions of protons and neutrons are also displayed.}
\label{fig:Pbee} 
\end{figure}
\begin{figure}[h]
\includegraphics[scale=0.8, angle=0]{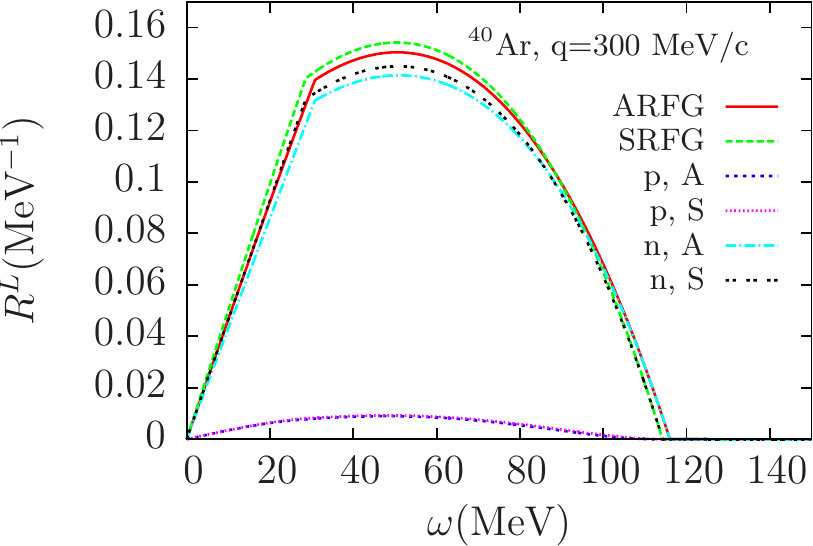}
\hspace{1.cm}
\includegraphics[scale=0.8, angle=0]{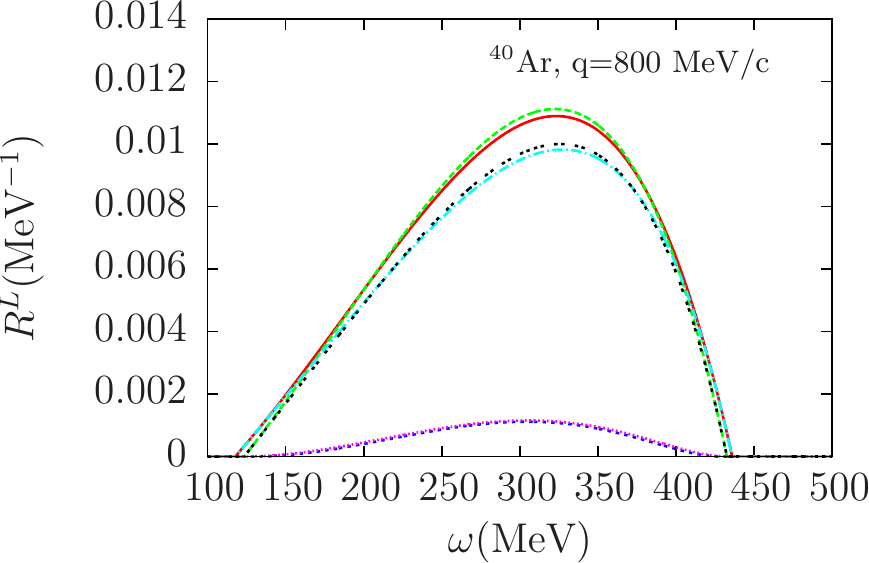}
\\
\includegraphics[scale=0.8, angle=0]{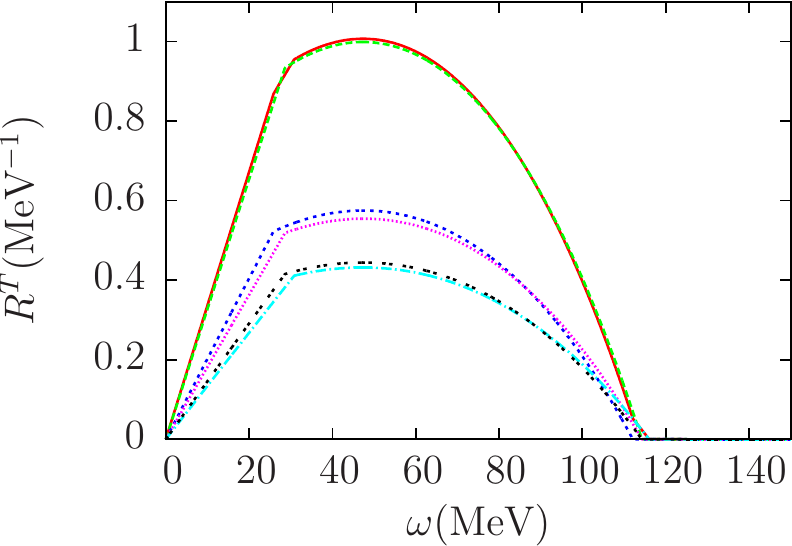}
\hspace{1.cm}
\includegraphics[scale=0.8, angle=0]{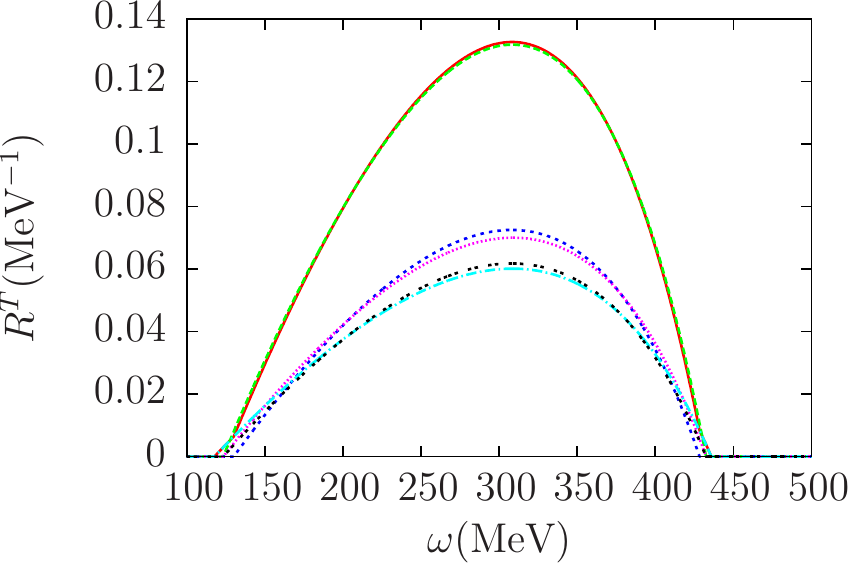}
\\
\includegraphics[scale=0.8, angle=0]{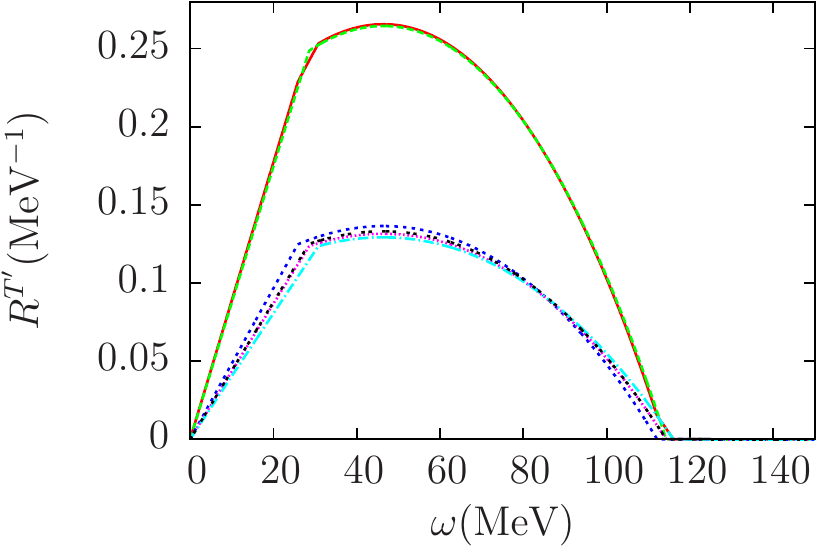}
\hspace{1.cm}
\includegraphics[scale=0.8, angle=0]{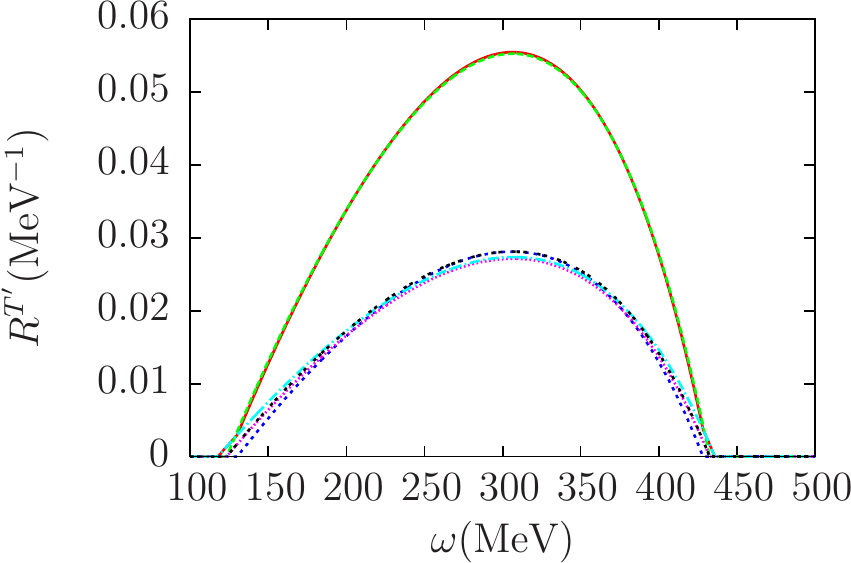}
\caption{(Color online) Weak neutral current response functions for $\nu$-$^{40}$Ar 
 in the symmetric (SRFG) and asymmetric (ARFG) relativistic Fermi gas.
 The separate contributions of protons and neutrons are also displayed.}
\label{fig:Arnunu} 
\end{figure}
\begin{figure}[h]
\includegraphics[scale=0.8, angle=0]{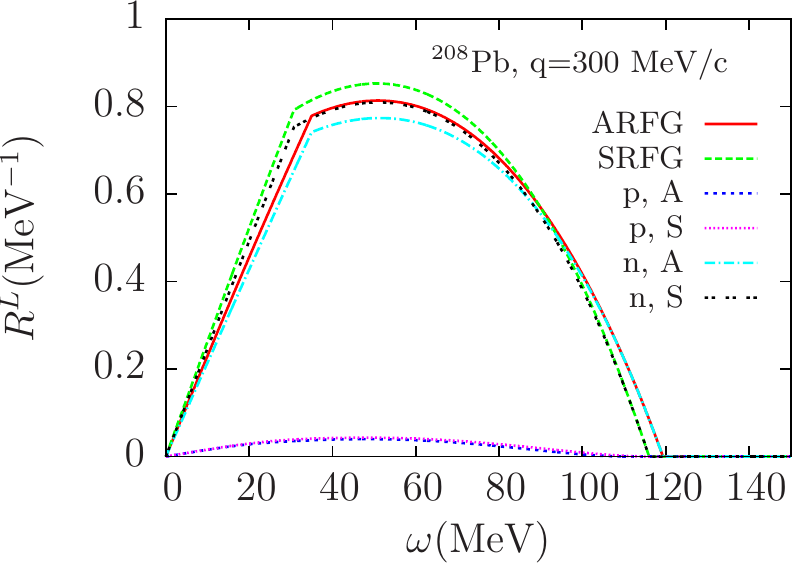}
\hspace{1.cm}
\includegraphics[scale=0.8, angle=0]{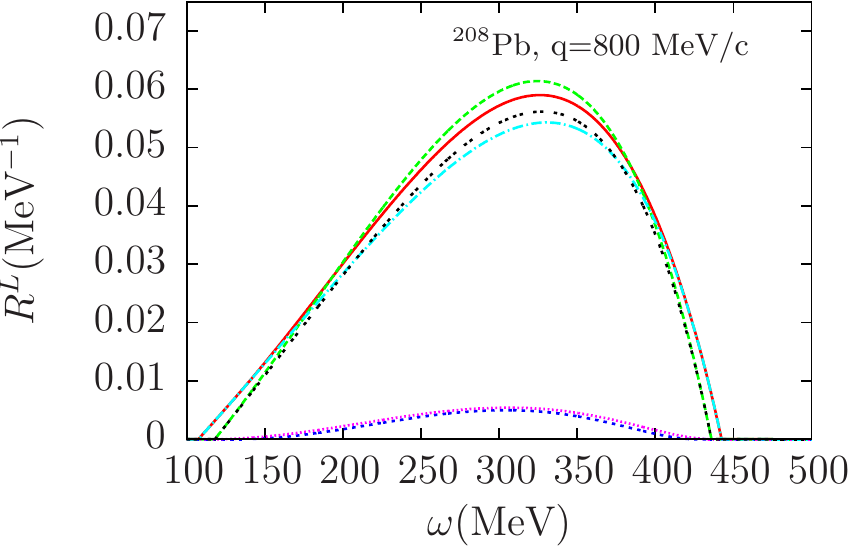}
\\
\includegraphics[scale=0.8, angle=0]{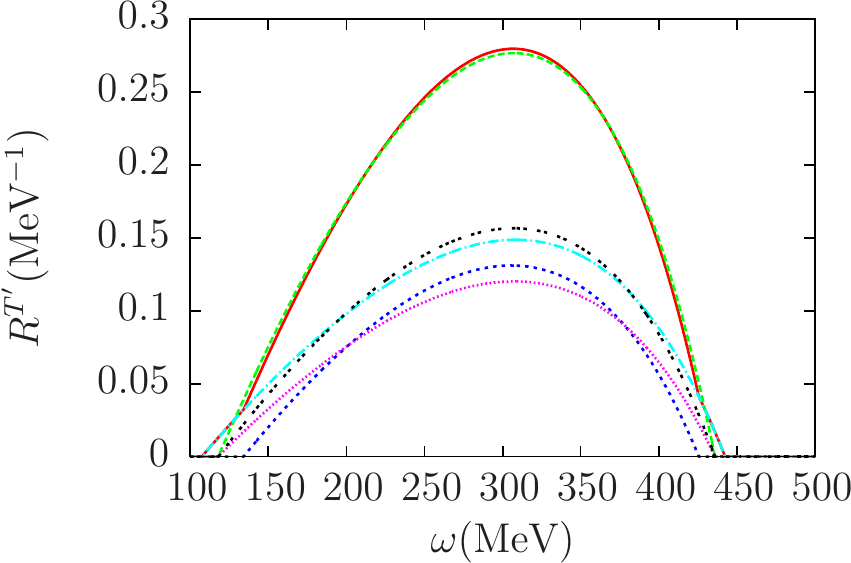}
\hspace{1.cm}
\includegraphics[scale=0.8, angle=0]{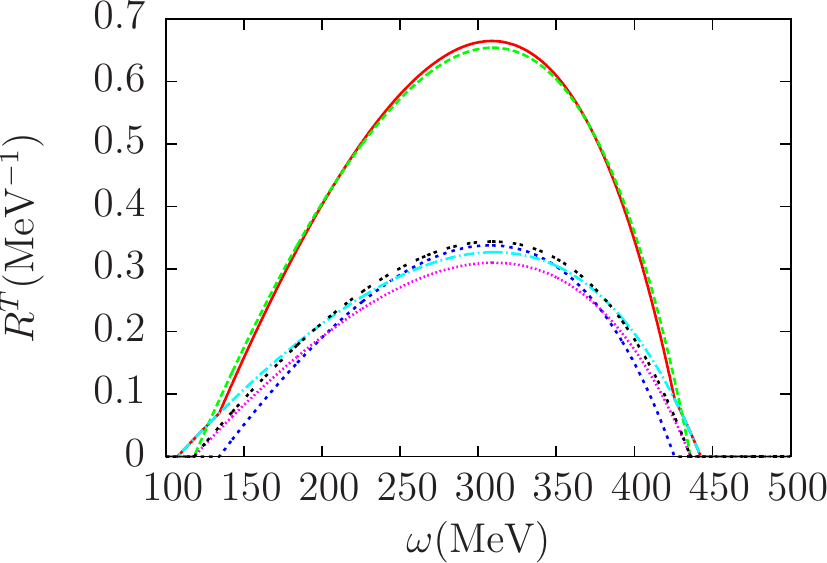}
\\
\includegraphics[scale=0.8, angle=0]{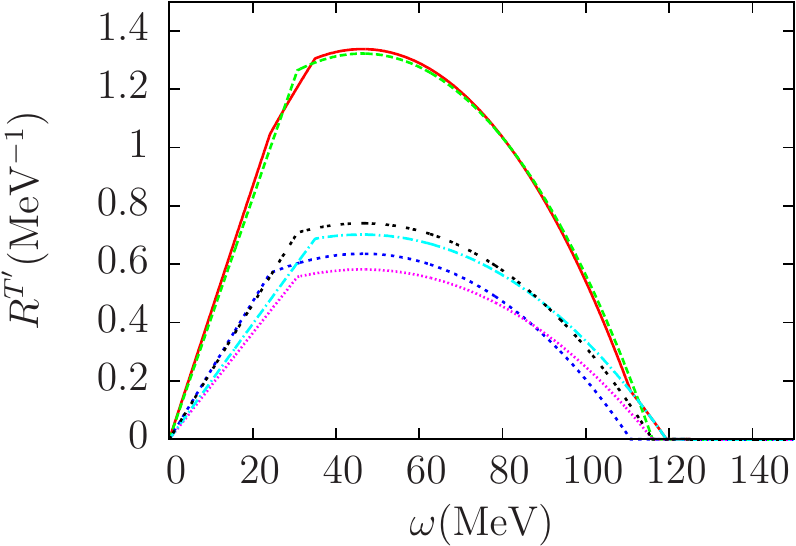}
\hspace{1.cm}
\includegraphics[scale=0.8, angle=0]{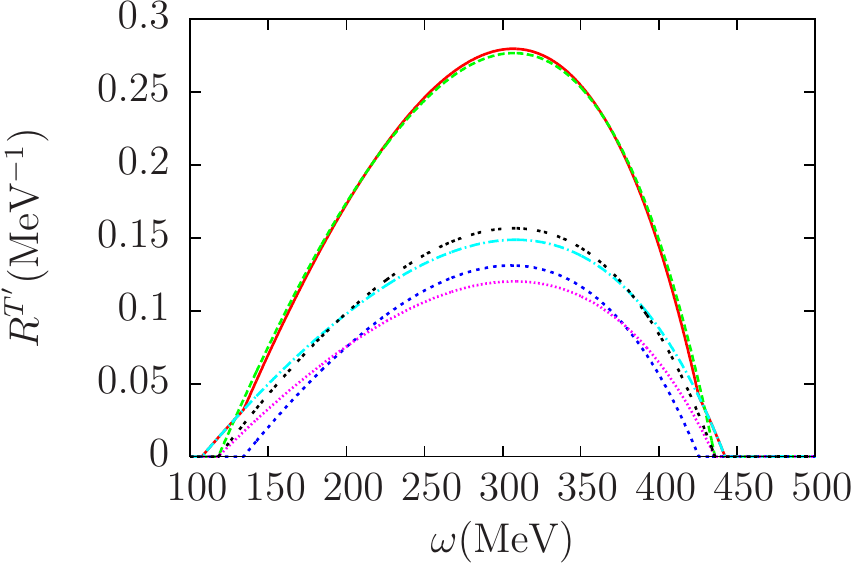}
\caption{(Color online) Weak neutral current response functions for $\nu$-$^{208}$Pb 
 in the symmetric (SRFG) and asymmetric (ARFG) relativistic Fermi gas.
 The separate contributions of protons and neutrons are also displayed.}
\label{fig:Pbnunu} 
\end{figure}
In Figs. \ref{fig:Aree} and \ref{fig:Pbee} we show the electromagnetic $(e,e')$ longitudinal and transverse response functions for $^{40}${Ar} and $^{208}$Pb as 
functions of the energy transfer $\omega$ and two values of the momentum transfers $q$. We also show the separate contributions of protons and neutrons. We observe that, as expected from the different values of the Fermi momentum, the proton ARFG responses are higher than the SRFG ones and limited to a narrower region of $\omega$, whereas the opposite occurs for neutrons. These two effects tend to cancel in the total transverse nuclear response $R^T$, which is affected only mildly by the N/Z asymmetry. On the other hand, in the longitudinal channel, $R^L$, where the proton response dominates, asymmetry effects can be non-negligible. Specifically, at $q=800$ MeV/c the ratios of the ARFG/SRFG responses at the quasielastic peak for argon are of the order of 1.03 (1.01) for $L(T)$ and are of the order of 1.07 (1.03) for $L(T)$ for lead. Roughly, the ratios are similar as functions of the momentum transfer.

In Figs. \ref{fig:Arnunu} and \ref{fig:Pbnunu} we show the weak neutral current longitudinal ($L$) and transverse (both $T$ and $T^\prime$) response functions for $^{40}${Ar} and $^{208}$Pb, obtained in the usual way \cite{Amaro:2004bs,PhysRevC.73.035503} by replacing the EM couplings by WNC couplings and adding the $T^\prime$ $VA$-interference response; recall that the last enters with the opposite sign for neutrinos and antineutrinos in the total cross section. Also, note that in this study we have ignored the effects from strangeness content in the nucleons. 

The effects are similar to what was found above for electron scattering, but not exactly the same for the cases which can be directly compared ({\it viz.,} $L$ and $T$), implying that for asymmetric nuclei there are effects to be taken into account in using input from electron scattering to obtain parts of the WNC cross section, as is often done in scaling analyses. The purpose of the present study is to get some idea about how significant such effects can be. Specifically, again at $q=800$ MeV/c the ratios of the ARFG/SRFG responses at the quasielastic peak for argon are of the order of 0.98 (1.01) for $L(T)$ and are of the order of 0.96 (1.02) for $L(T)$ for lead, while the ratios for the $T^\prime$ response are of the order of 1.00 and 1.01 for argon and lead, respectively. Furthermore, in $(\nu,\nu')$ the ARFG responses are lower and more extended than the SRFG ones, while the opposite occurs for  $(\bar\nu,\bar\nu')$.

\subsection{Charged-current reactions}
\label{subsec:resCC}

Let us now consider CC neutrino and antineutrino reactions. In this case the inclusive cross section is the combination of 5, instead of 3, response functions, as a consequence of the non-conservation of the axial current and of the non-vanishing mass of the outgoing charged lepton.

\begin{figure}[h]
  \includegraphics[scale=0.8, angle=0]{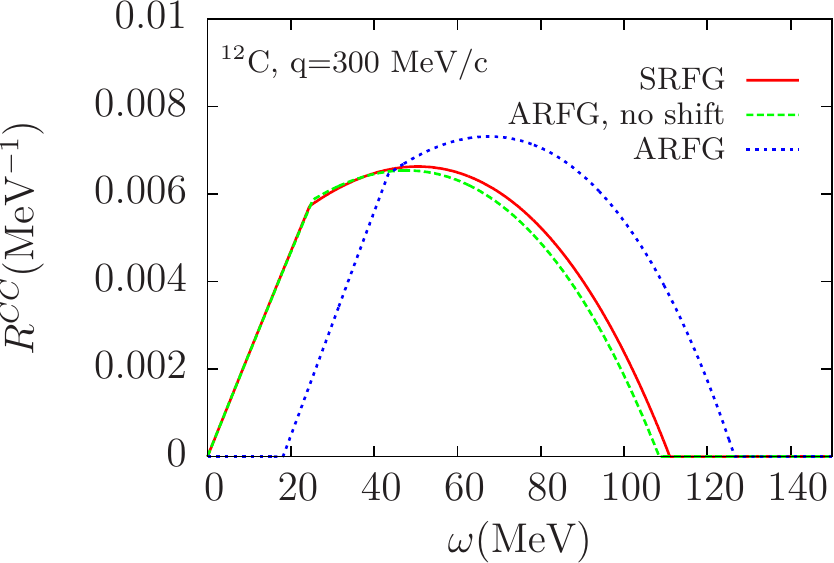}
\hspace{1.cm}
\includegraphics[scale=0.8, angle=0]{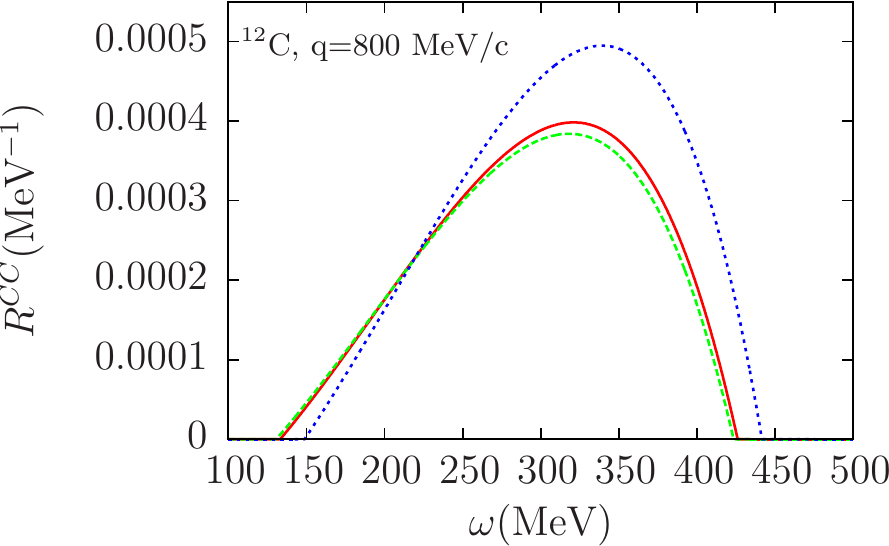}
\\
\includegraphics[scale=0.8, angle=0]{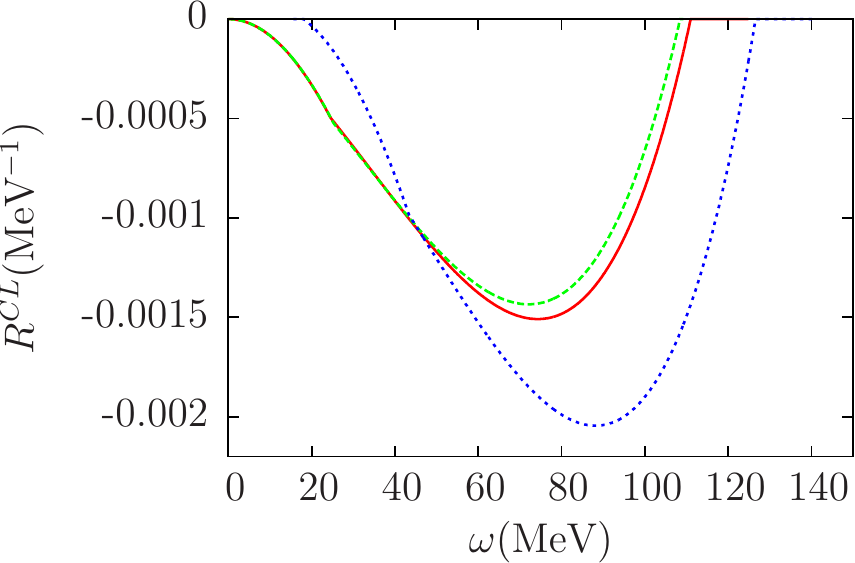}
\hspace{1.cm}
\includegraphics[scale=0.8, angle=0]{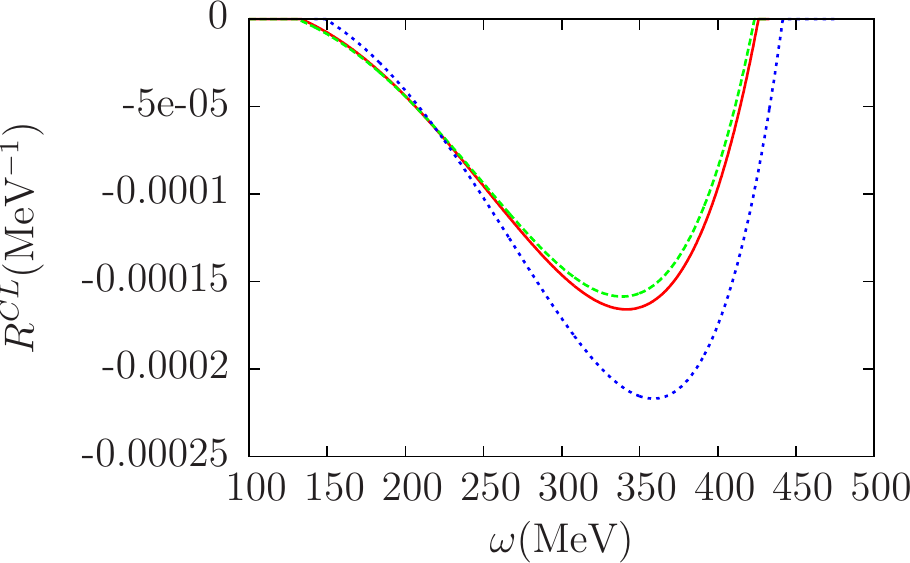}
\\
\includegraphics[scale=0.8, angle=0]{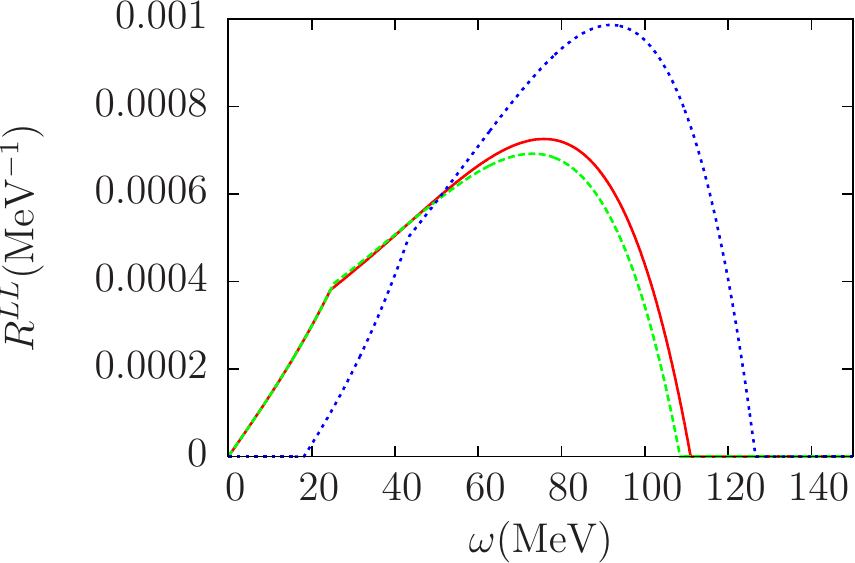}
\hspace{1.cm}
\includegraphics[scale=0.8, angle=0]{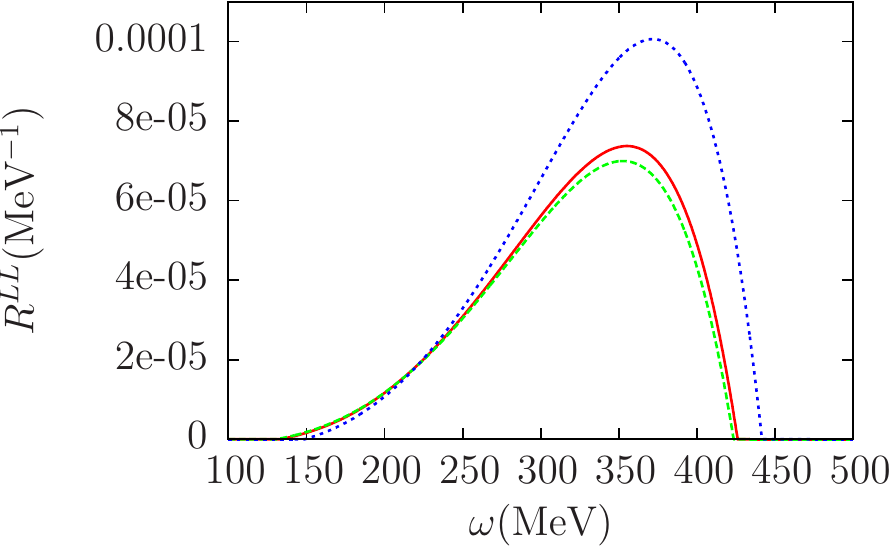}
\\
\includegraphics[scale=0.8, angle=0]{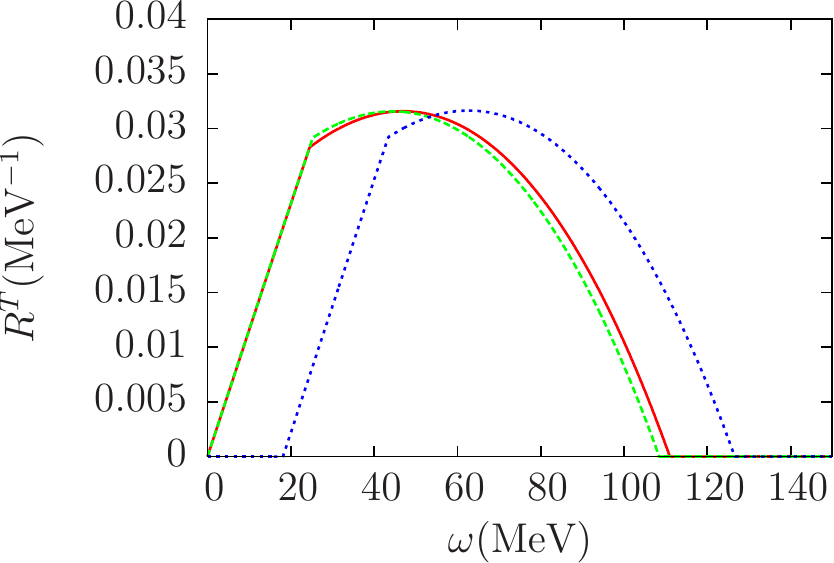}
\hspace{1.cm}
\includegraphics[scale=0.8, angle=0]{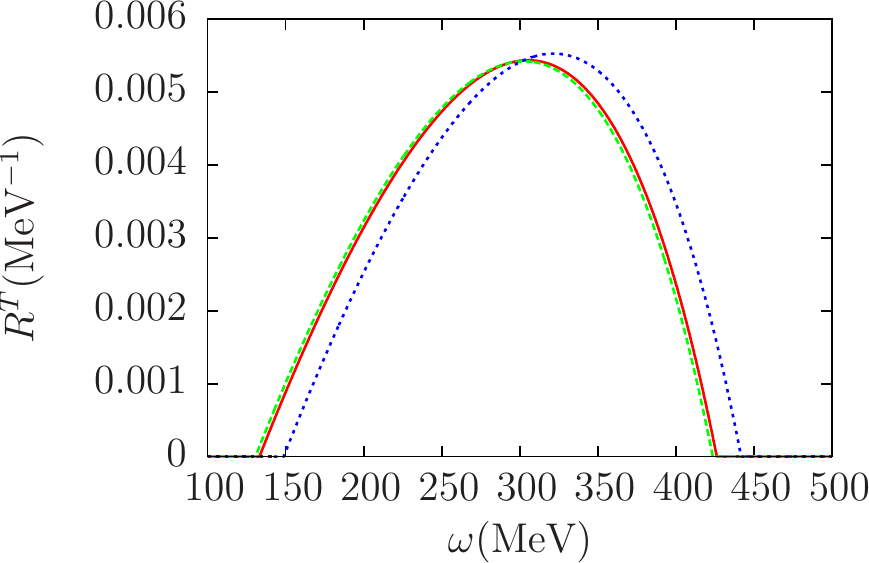}
\\
\includegraphics[scale=0.8, angle=0]{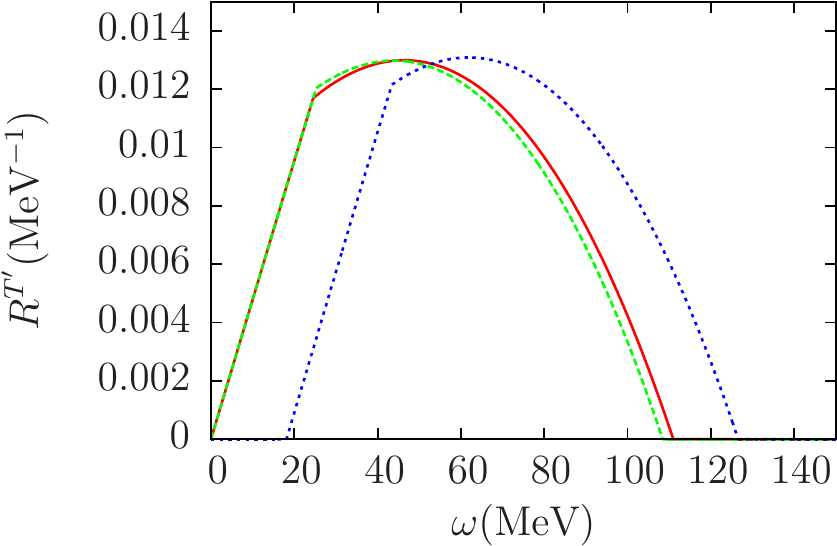}
\hspace{1.cm}
\includegraphics[scale=0.8, angle=0]{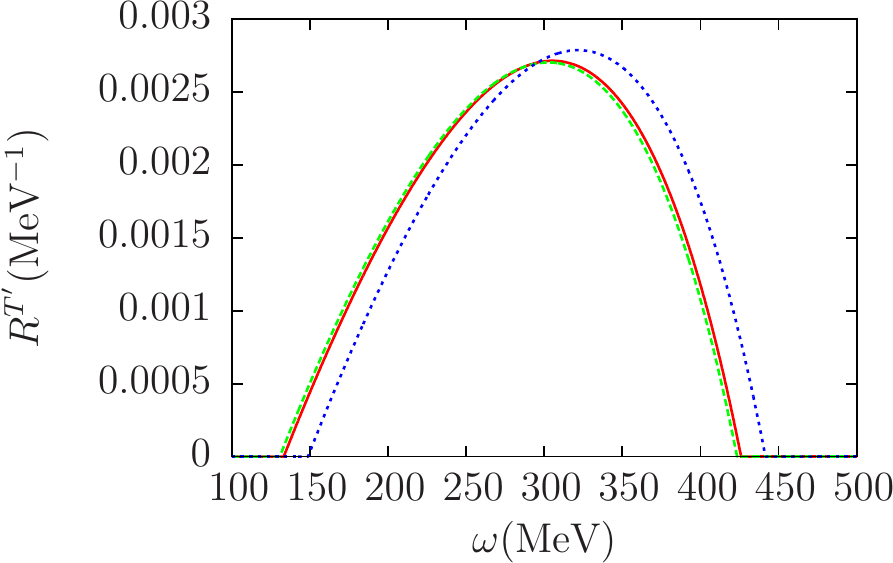}
\caption{(Color online) Neutrino charged-current weak response functions per neutron of $^{12}$C 
 in the symmetric (SRFG) and asymmetric (ARFG) relativistic Fermi gas. The ARFG results with no energy shift (see text) are also shown.
 Each column corresponds to a fixed value of the momentum transfer $q$.}
\label{fig:Cnumu} 
\end{figure}

\begin{figure}[h]
  \includegraphics[scale=0.8, angle=0]{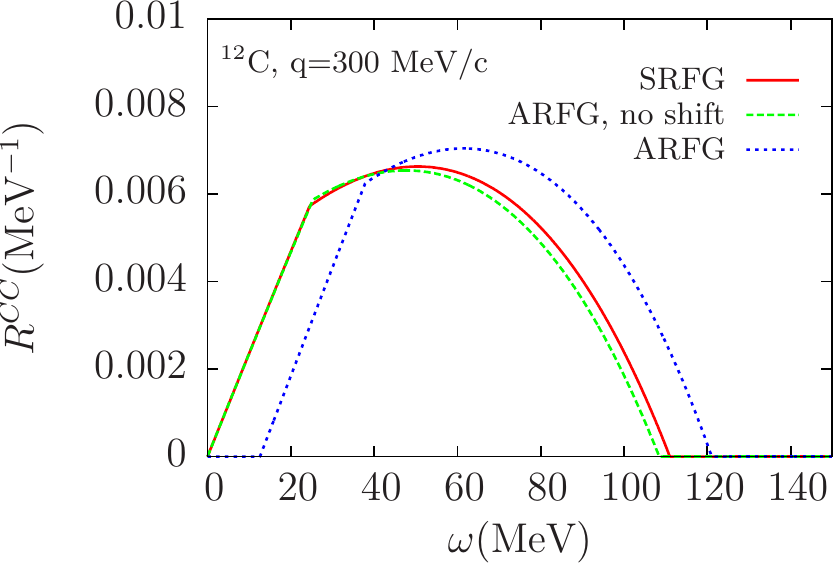}
\hspace{1.cm}
\includegraphics[scale=0.8, angle=0]{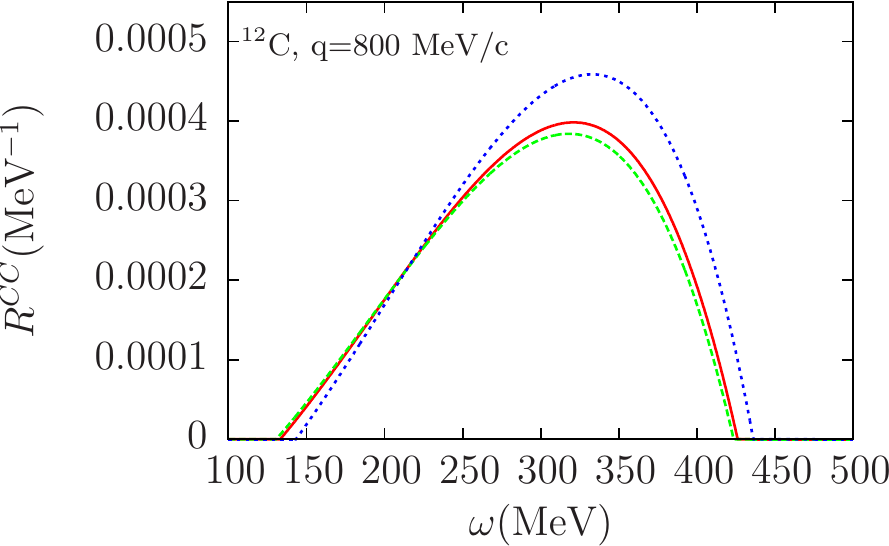}
\\
\includegraphics[scale=0.8, angle=0]{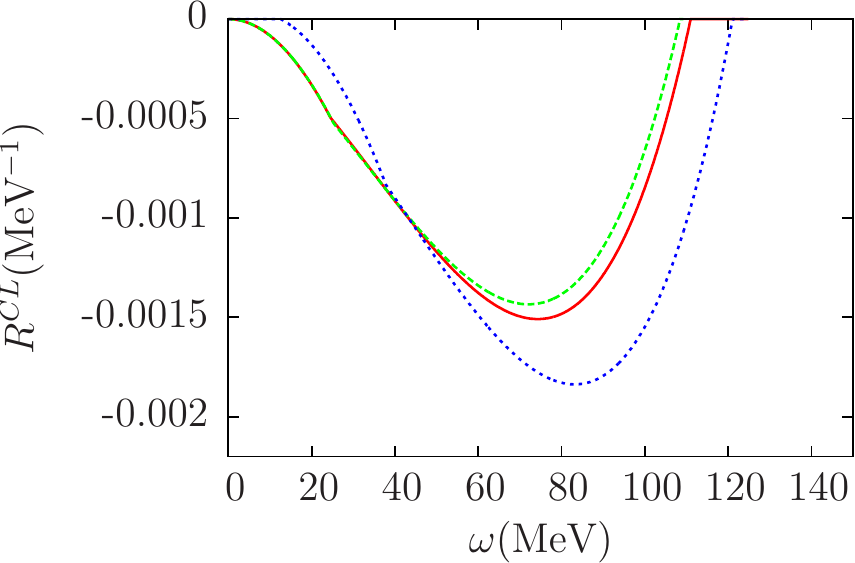}
\hspace{1.cm}
\includegraphics[scale=0.8, angle=0]{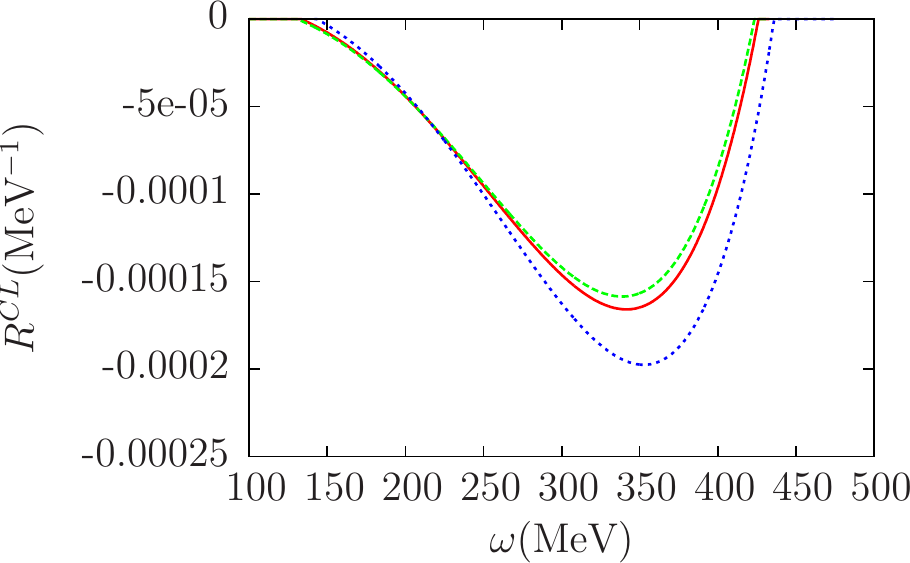}
\\
\includegraphics[scale=0.8, angle=0]{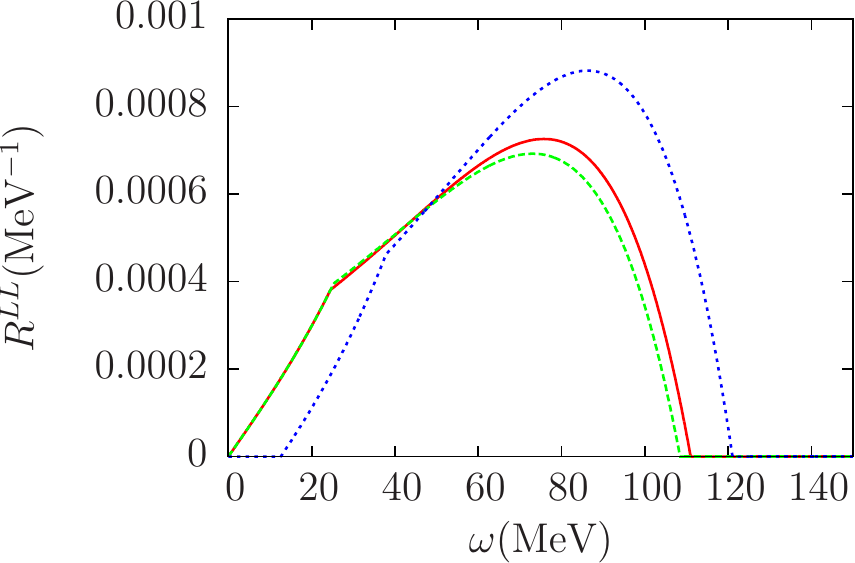}
\hspace{1.cm}
\includegraphics[scale=0.8, angle=0]{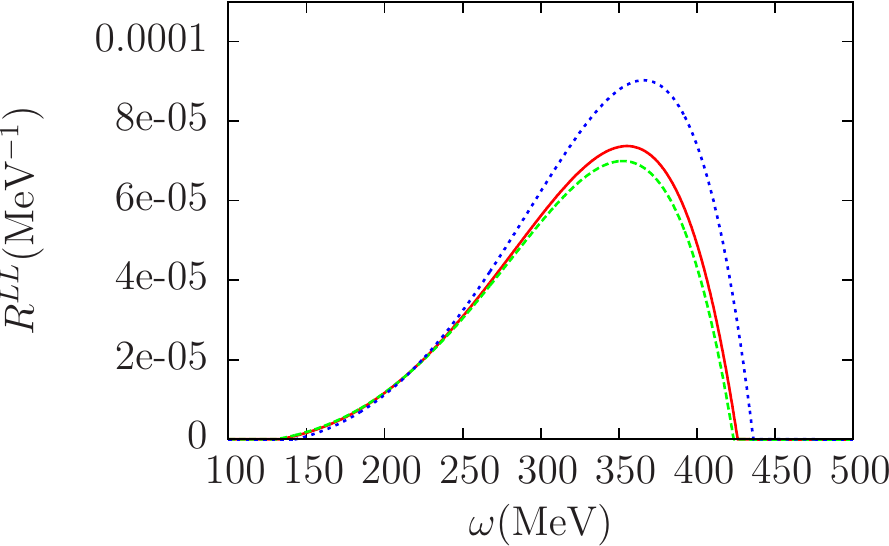}
\\
\includegraphics[scale=0.8, angle=0]{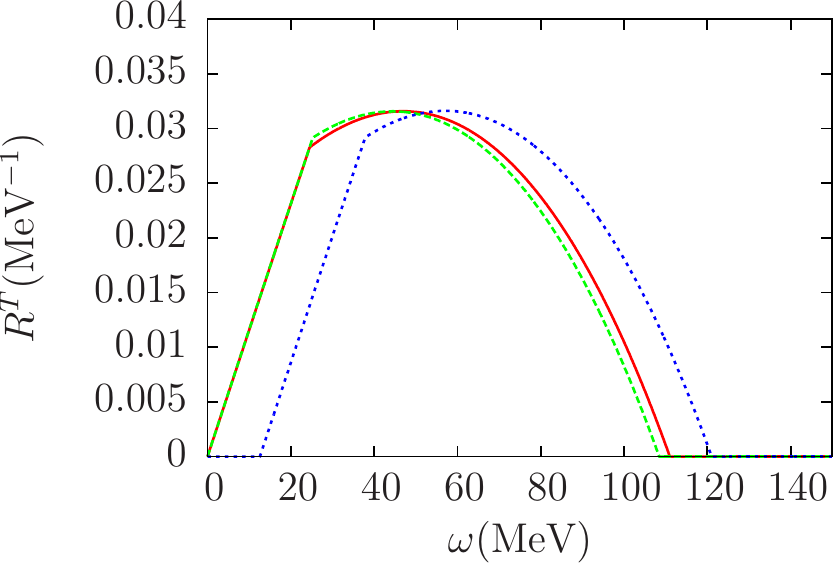}
\hspace{1.cm}
\includegraphics[scale=0.8, angle=0]{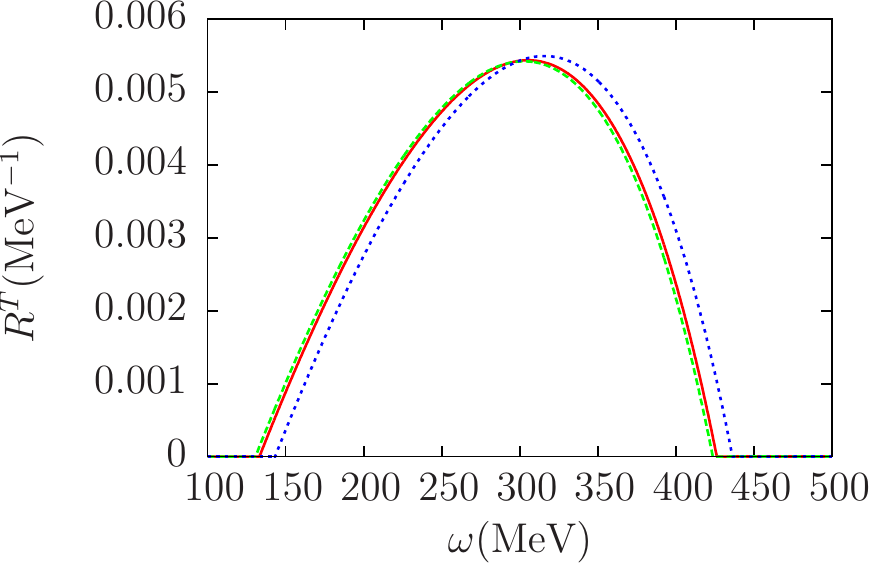}
\\
\includegraphics[scale=0.8, angle=0]{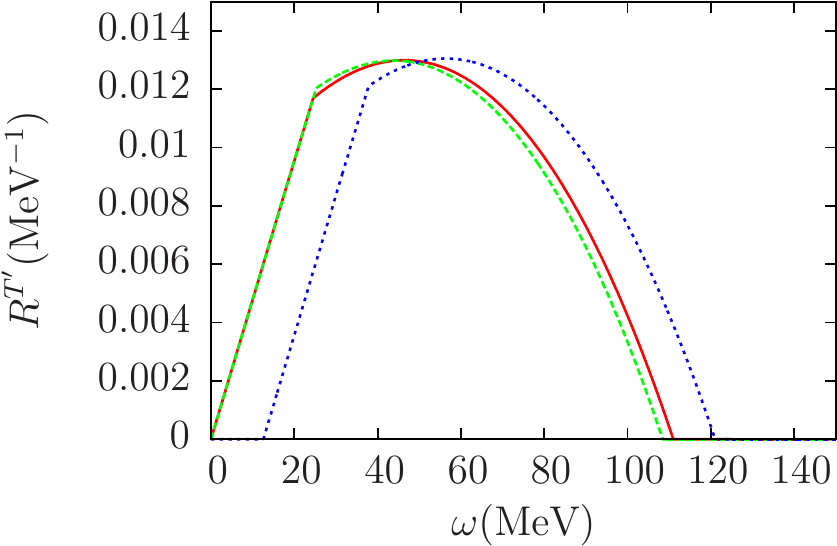}
\hspace{1.cm}
\includegraphics[scale=0.8, angle=0]{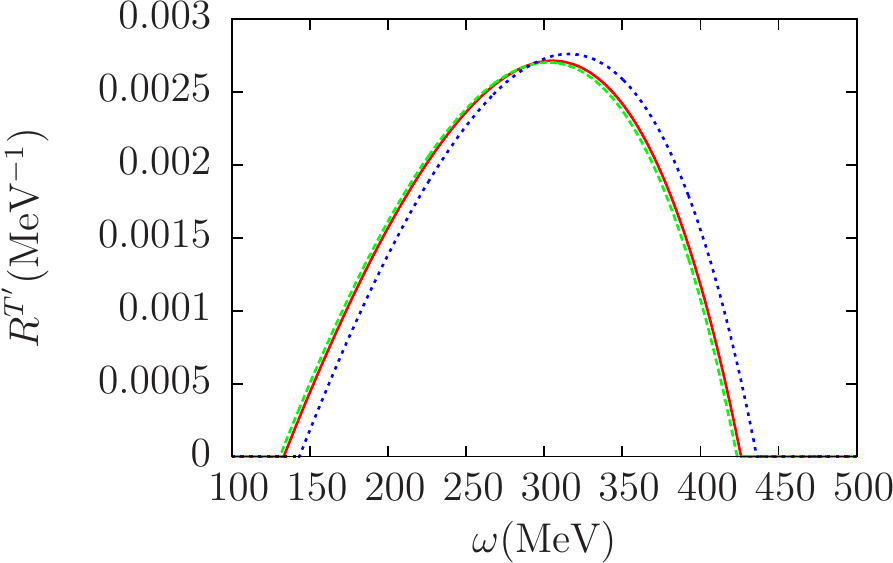}
\caption{(Color online) Antineutrino charged-current weak response functions per proton of $^{12}$C 
 in the symmetric (SRFG) and asymmetric (ARFG) relativistic Fermi gas. The ARFG results with no energy shift (see text) are also shown.
 Each column corresponds to a fixed value of the momentum transfer $q$.}
\label{fig:Cnubarmu} 
\end{figure}

\begin{figure}[h]
  \includegraphics[scale=0.8, angle=0]{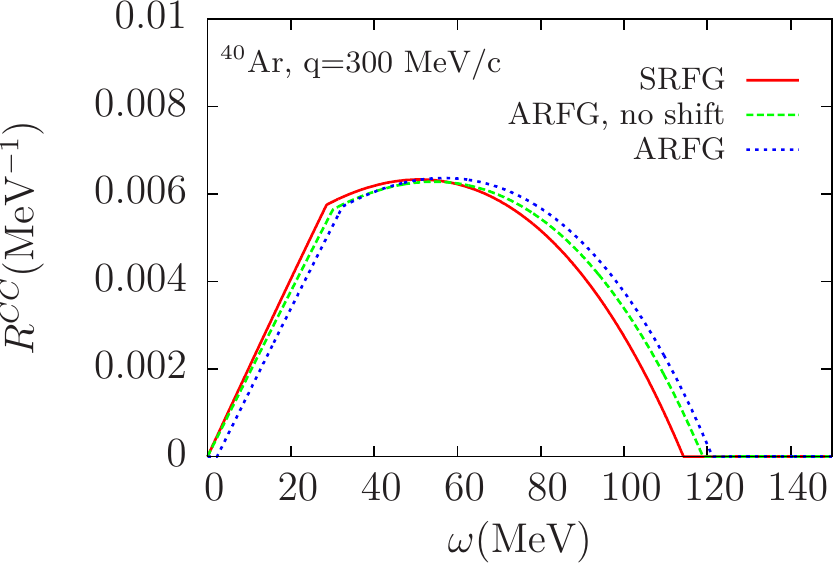}
\hspace{1.cm}
\includegraphics[scale=0.8, angle=0]{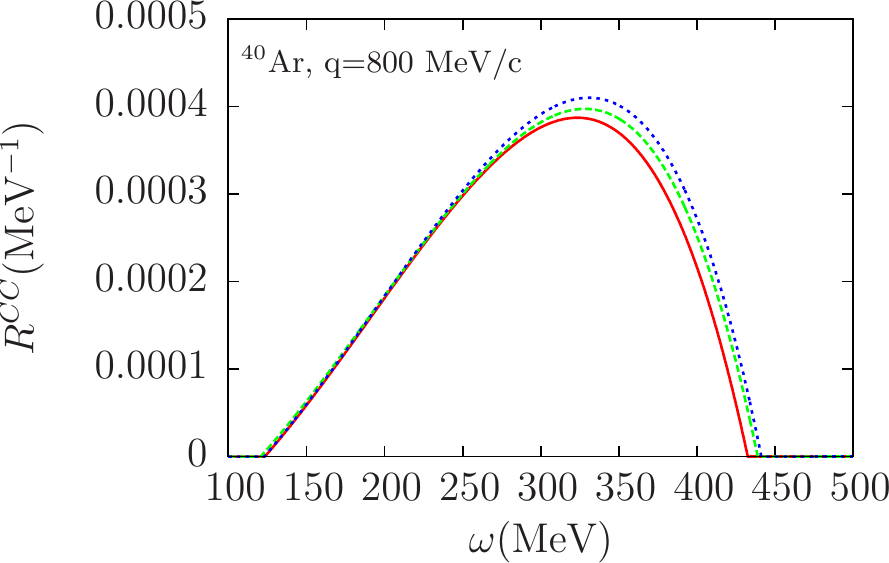}
\\
\includegraphics[scale=0.8, angle=0]{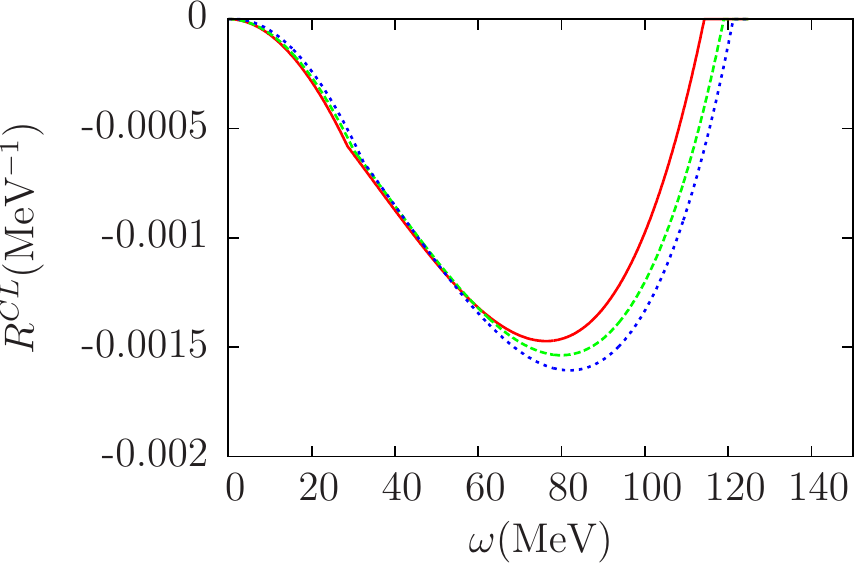}
\hspace{1.cm}
\includegraphics[scale=0.8, angle=0]{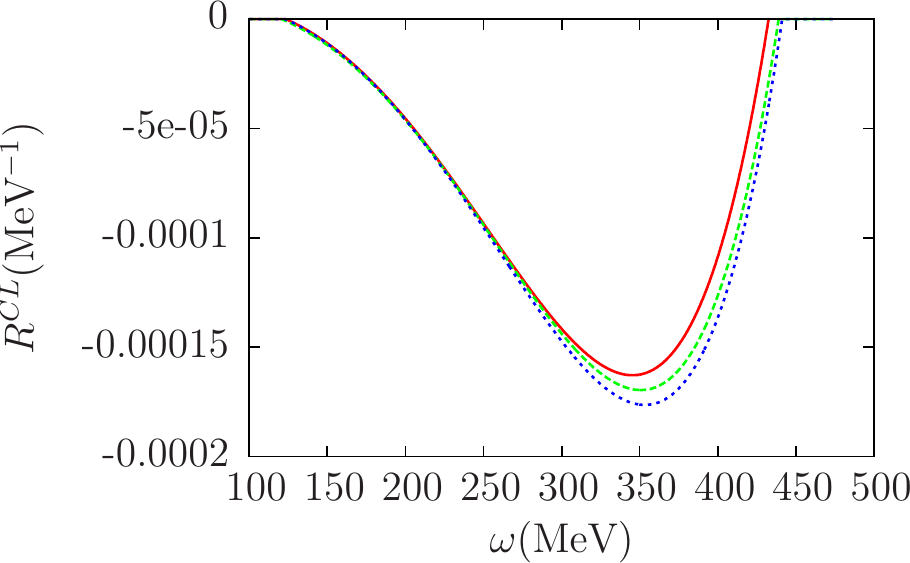}
\\
\includegraphics[scale=0.8, angle=0]{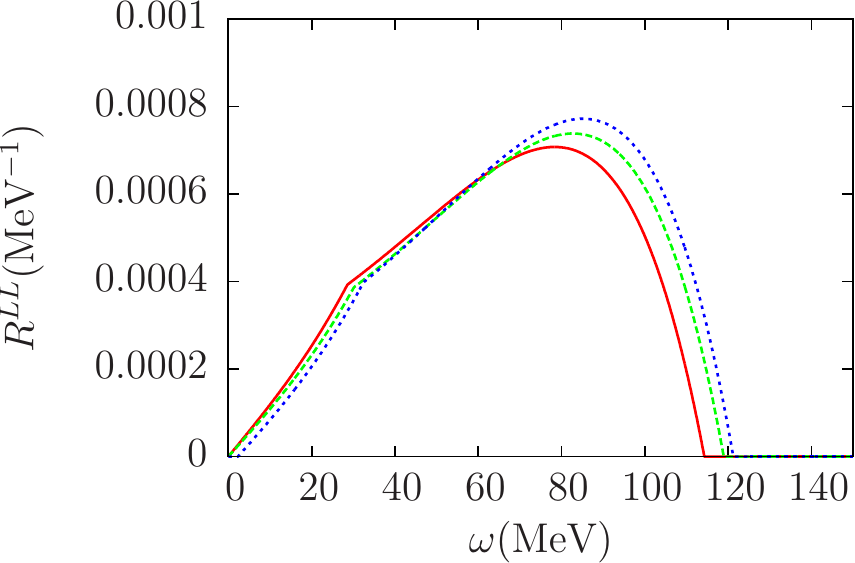}
\hspace{1.cm}
\includegraphics[scale=0.8, angle=0]{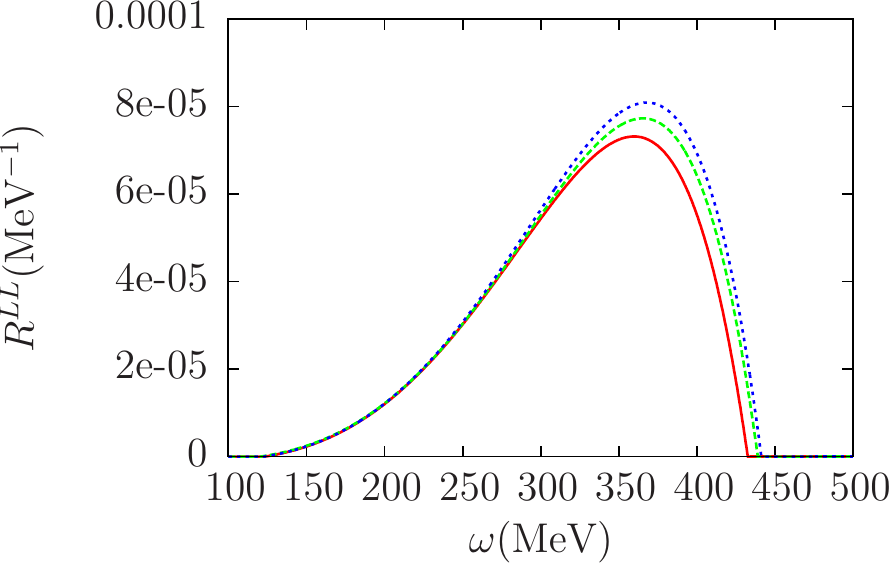}
\\
\includegraphics[scale=0.8, angle=0]{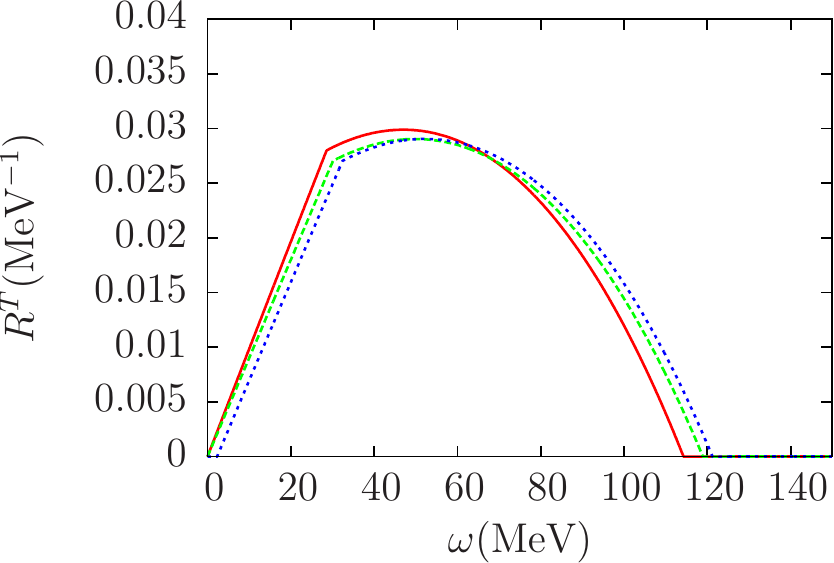}
\hspace{1.cm}
\includegraphics[scale=0.8, angle=0]{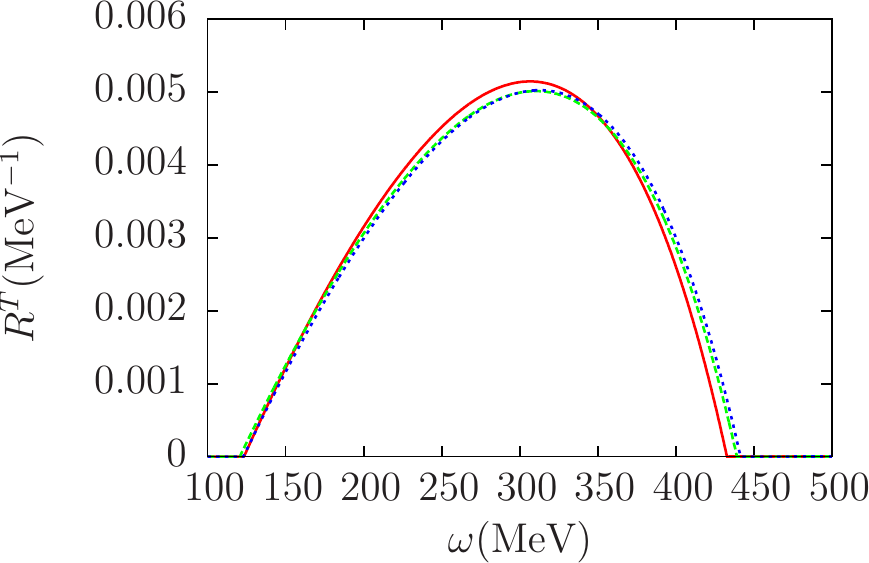}
\\
\includegraphics[scale=0.8, angle=0]{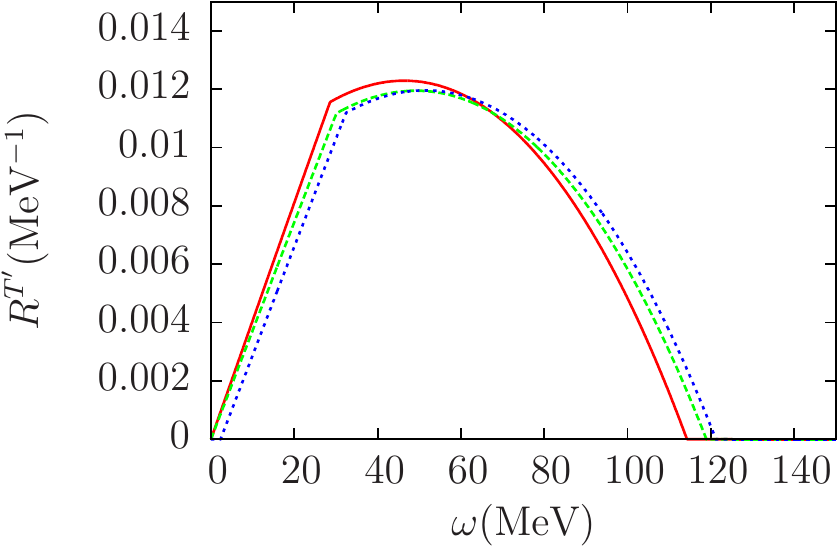}
\hspace{1.cm}
\includegraphics[scale=0.8, angle=0]{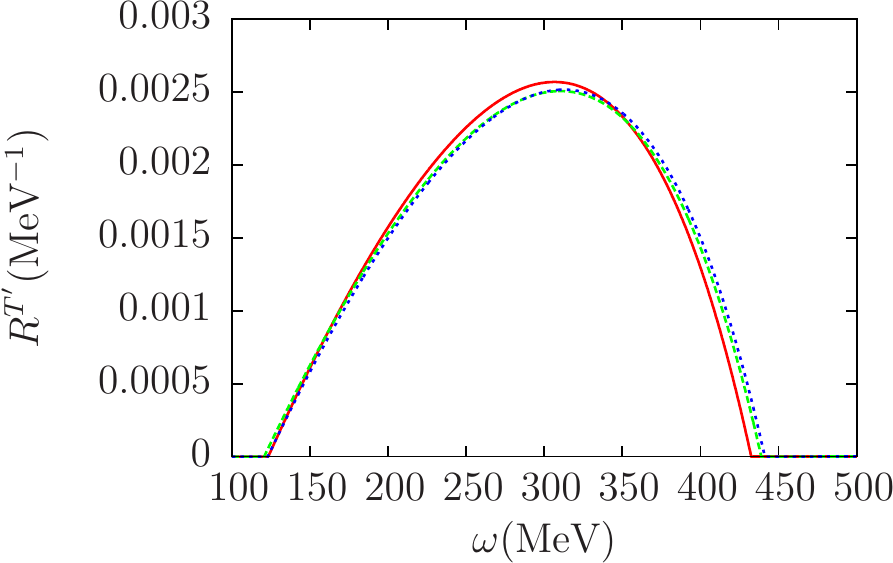}
\caption{(Color online) Neutrino charged-current weak response functions per neutron of $^{40}$Ar 
 in the symmetric (SRFG) and asymmetric (ARFG) relativistic Fermi gas. The ARFG results with no energy shift (see text) are also shown.
 Each column corresponds to a fixed value of the momentum transfer $q$.}
\label{fig:Arnumu} 
\end{figure}

\begin{figure}[h]
  \includegraphics[scale=0.8, angle=0]{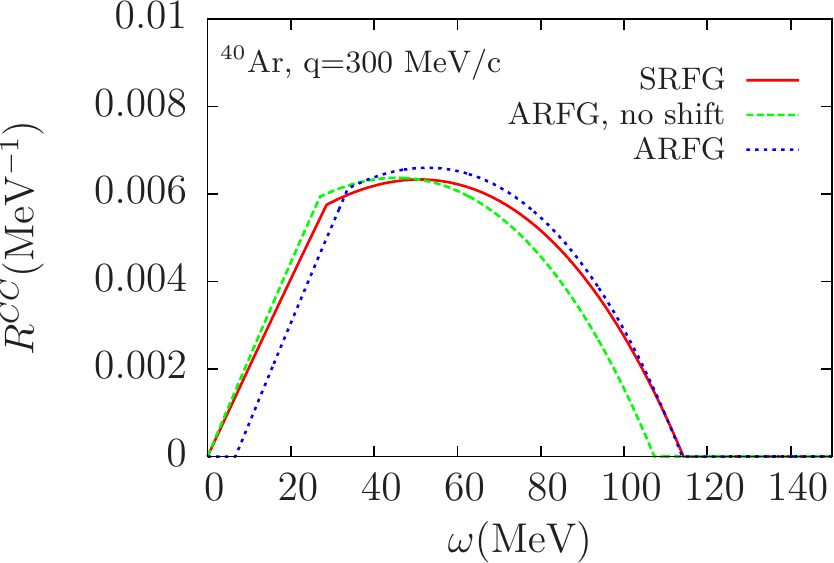}
\hspace{1.cm}
\includegraphics[scale=0.8, angle=0]{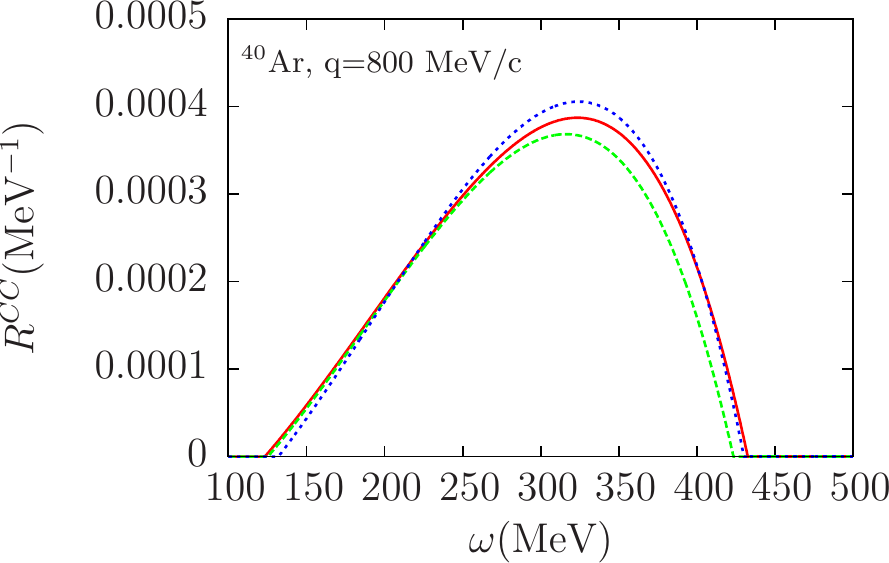}
\\
\includegraphics[scale=0.8, angle=0]{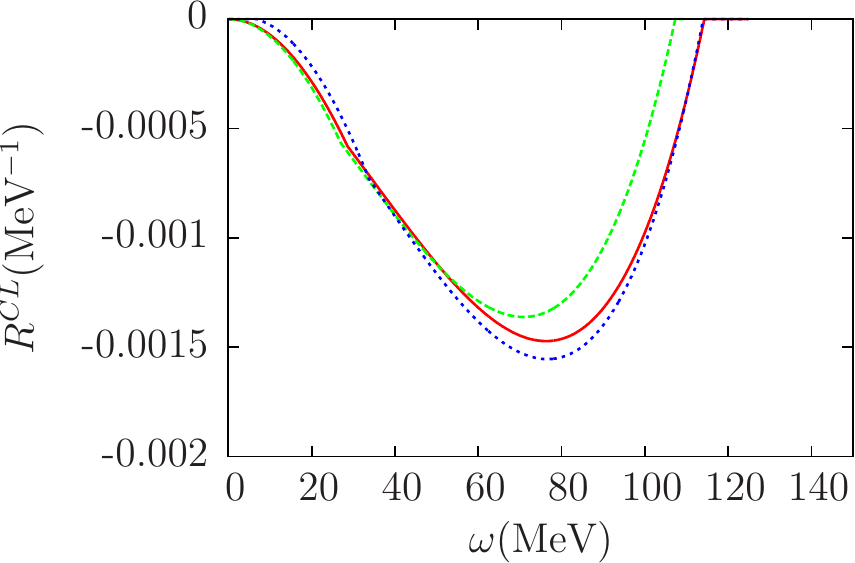}
\hspace{1.cm}
\includegraphics[scale=0.8, angle=0]{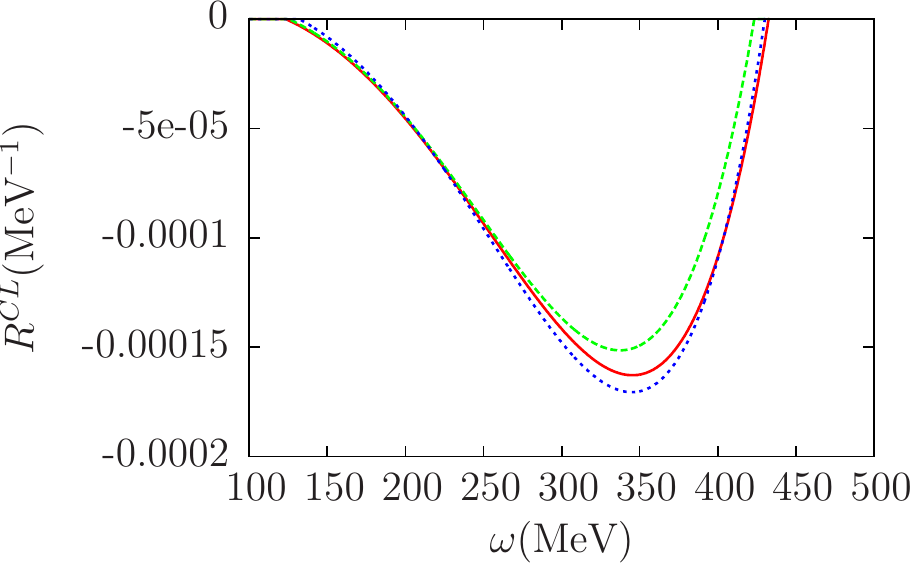}
\\
\includegraphics[scale=0.8, angle=0]{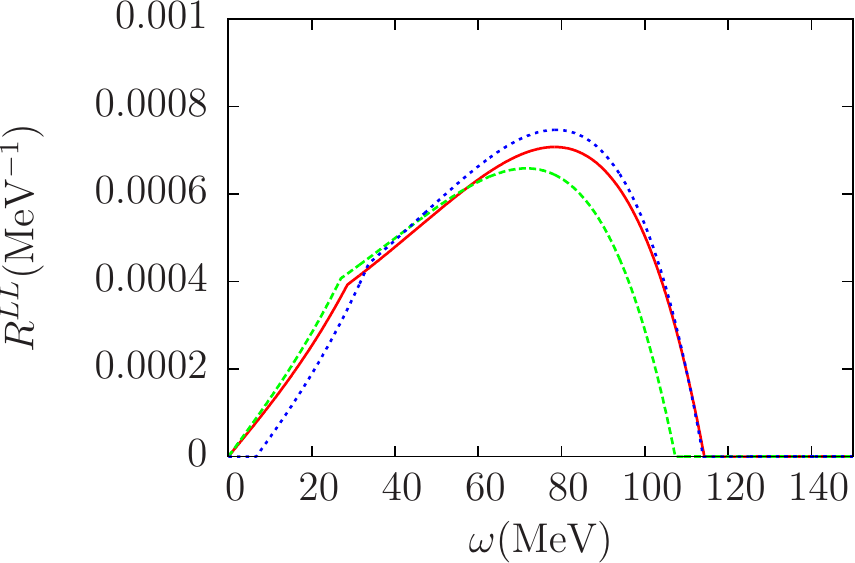}
\hspace{1.cm}
\includegraphics[scale=0.8, angle=0]{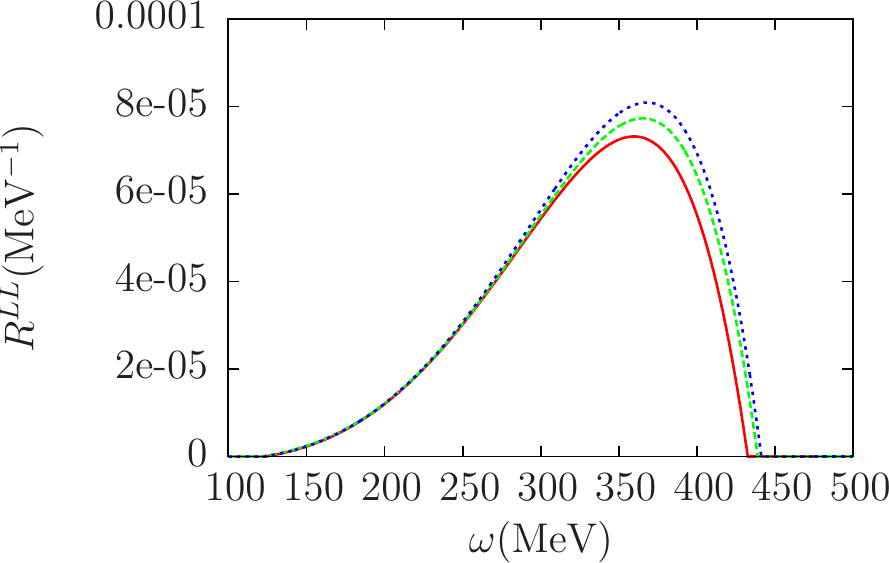}
\\
\includegraphics[scale=0.8, angle=0]{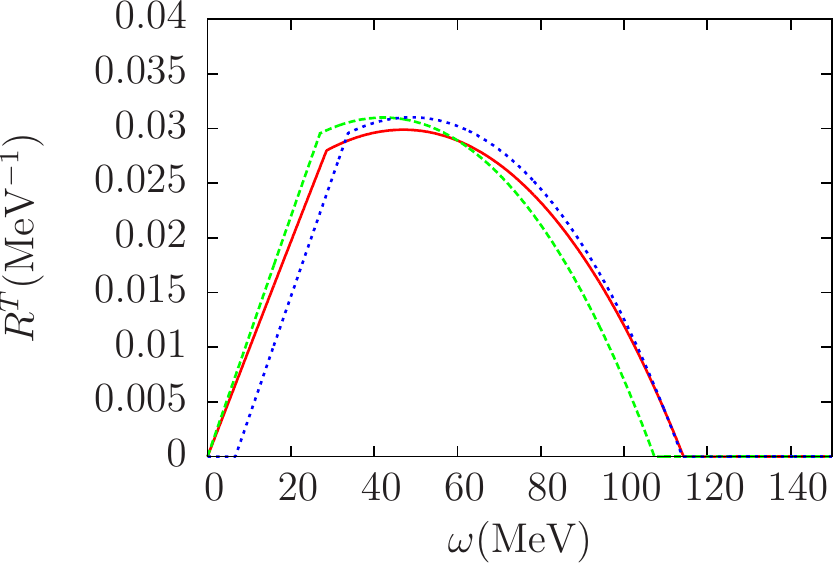}
\hspace{1.cm}
\includegraphics[scale=0.8, angle=0]{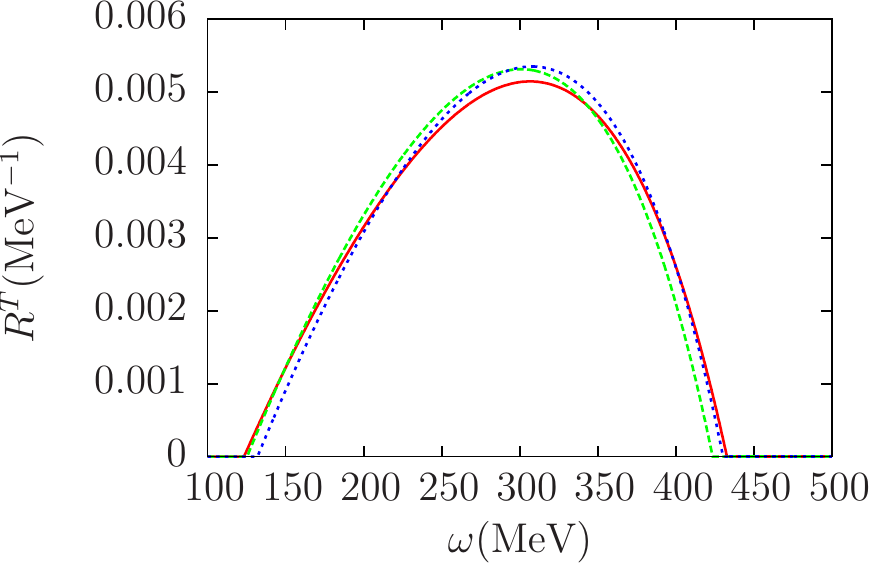}
\\
\includegraphics[scale=0.8, angle=0]{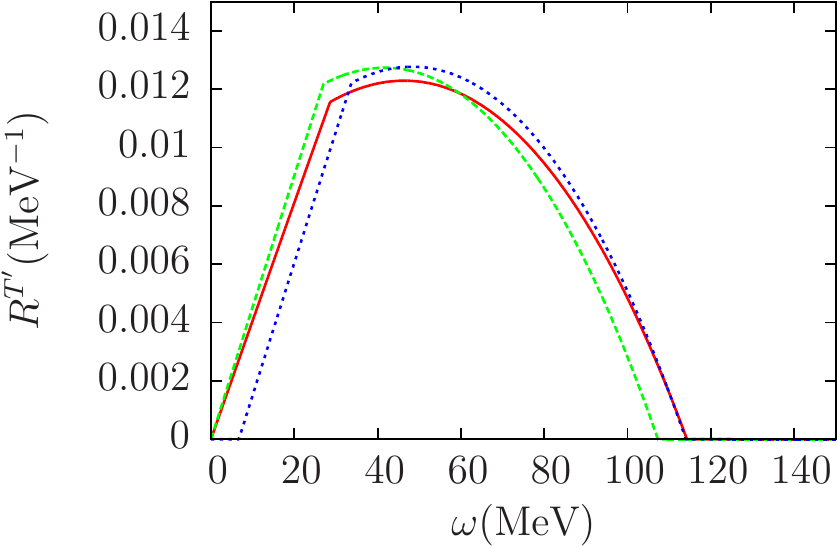}
\hspace{1.cm}
\includegraphics[scale=0.8, angle=0]{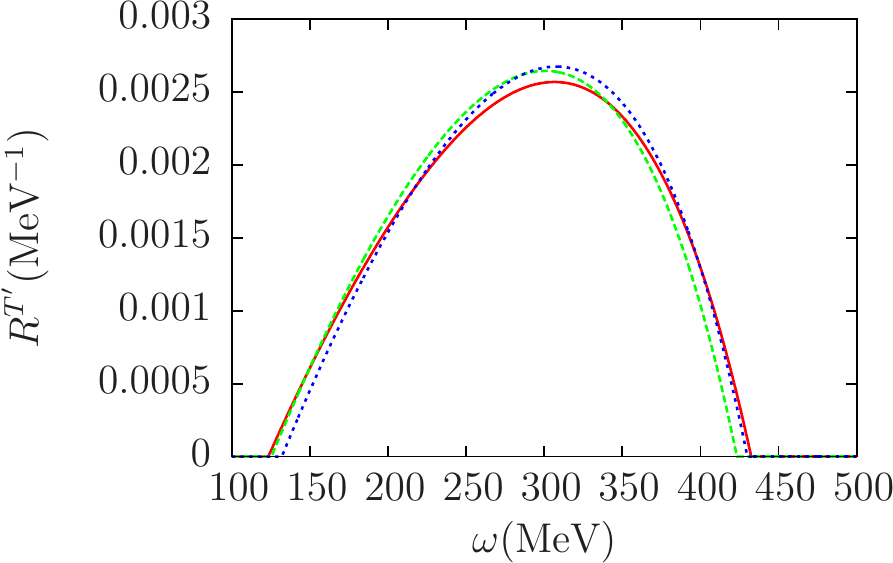}
\caption{(Color online) Antineutrino charged-current weak response functions per proton of $^{40}$Ar 
 in the symmetric (SRFG) and asymmetric (ARFG) relativistic Fermi gas. The ARFG results with no energy shift (see text) are also shown.
 Each column corresponds to a fixed value of the momentum transfer $q$.}
\label{fig:Arnubarmu} 
\end{figure}

\begin{figure}[h]
  \includegraphics[scale=0.8, angle=0]{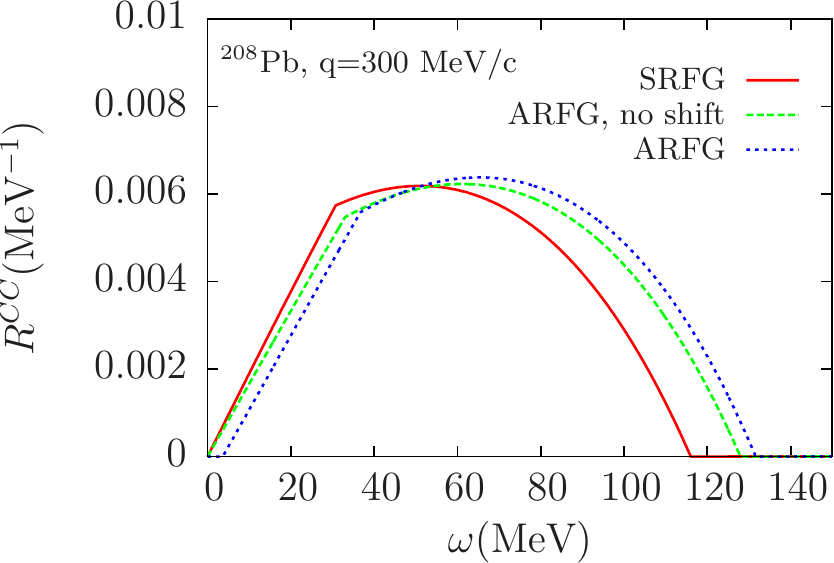}
\hspace{1.cm}
\includegraphics[scale=0.8, angle=0]{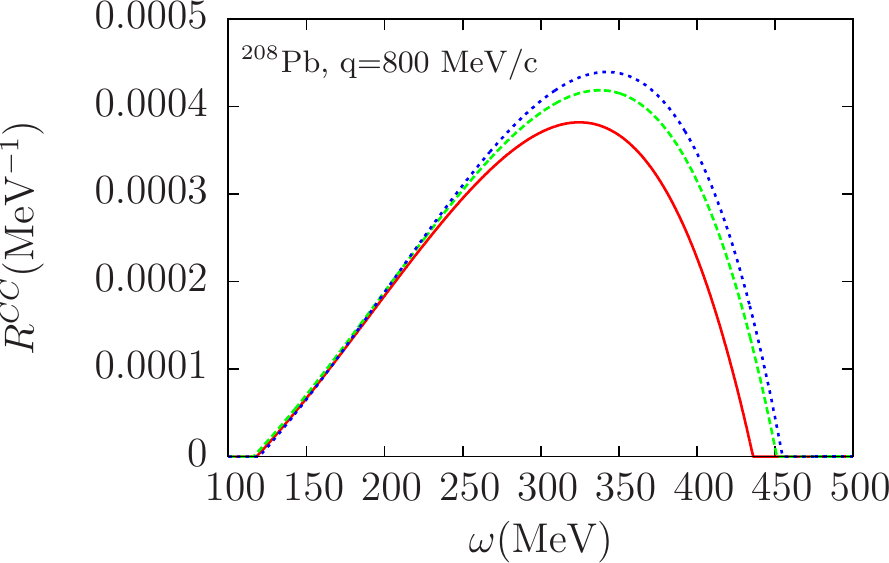}
\\
\includegraphics[scale=0.8, angle=0]{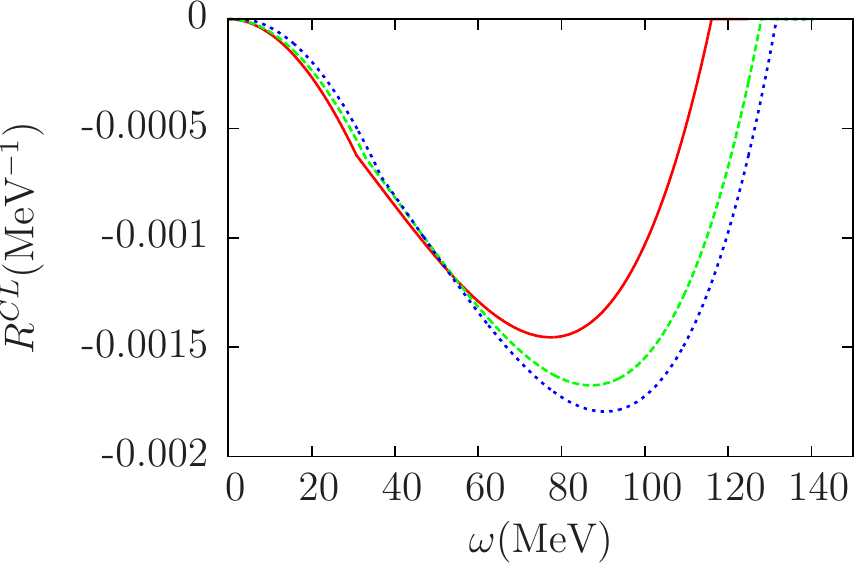}
\hspace{1.cm}
\includegraphics[scale=0.8, angle=0]{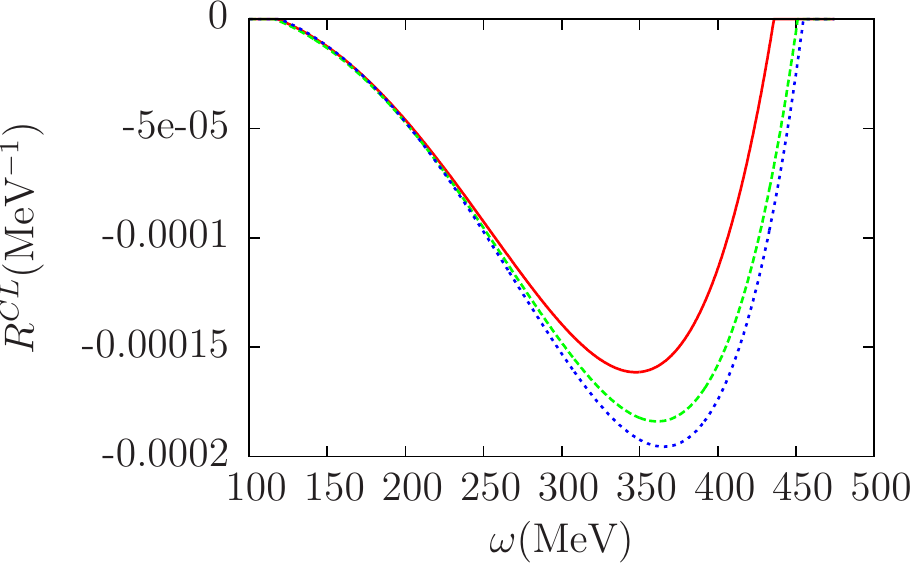}
\\
\includegraphics[scale=0.8, angle=0]{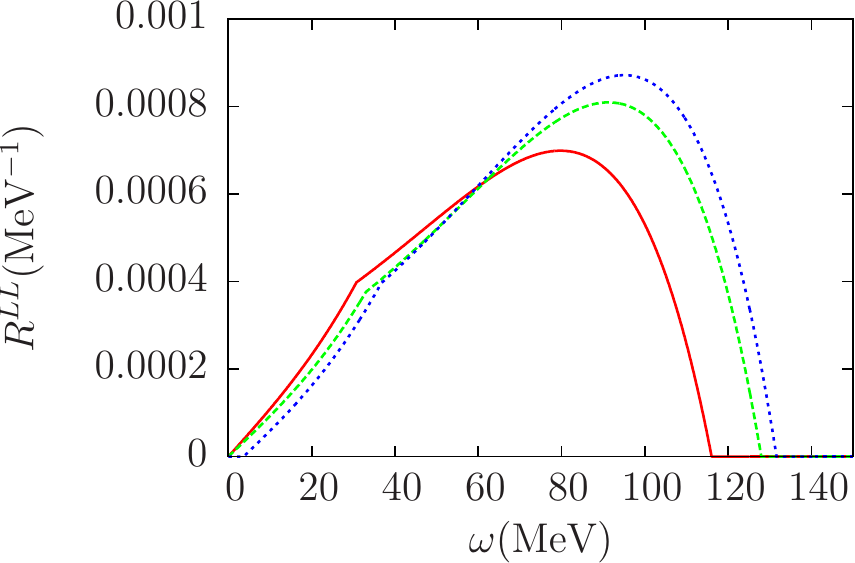}
\hspace{1.cm}
\includegraphics[scale=0.8, angle=0]{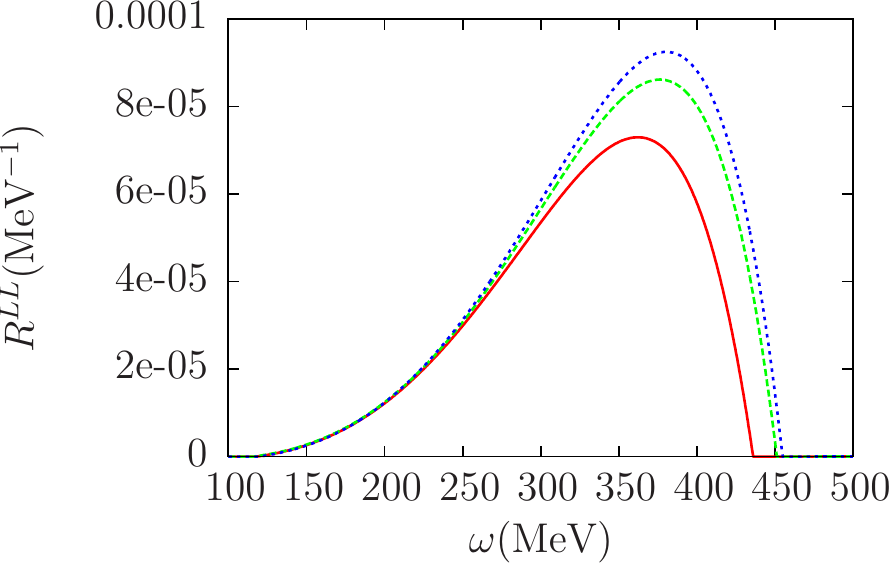}
\\
\includegraphics[scale=0.8, angle=0]{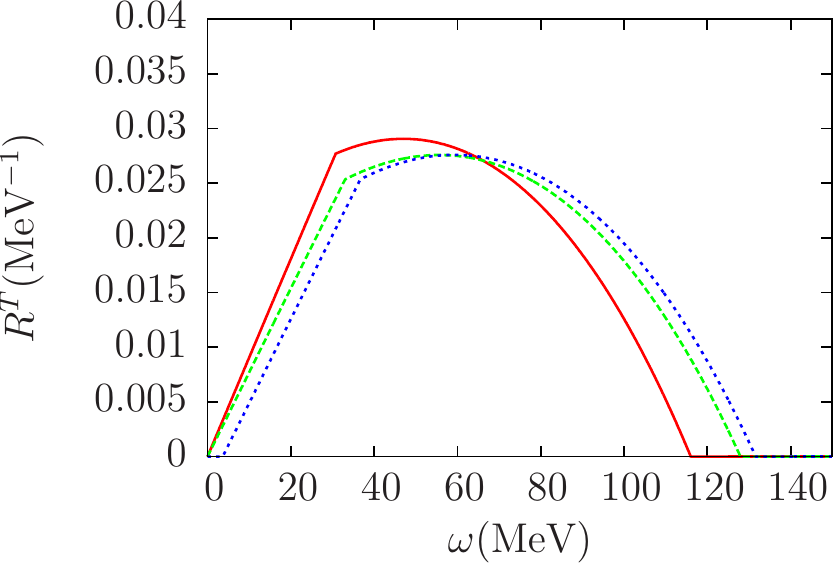}
\hspace{1.cm}
\includegraphics[scale=0.8, angle=0]{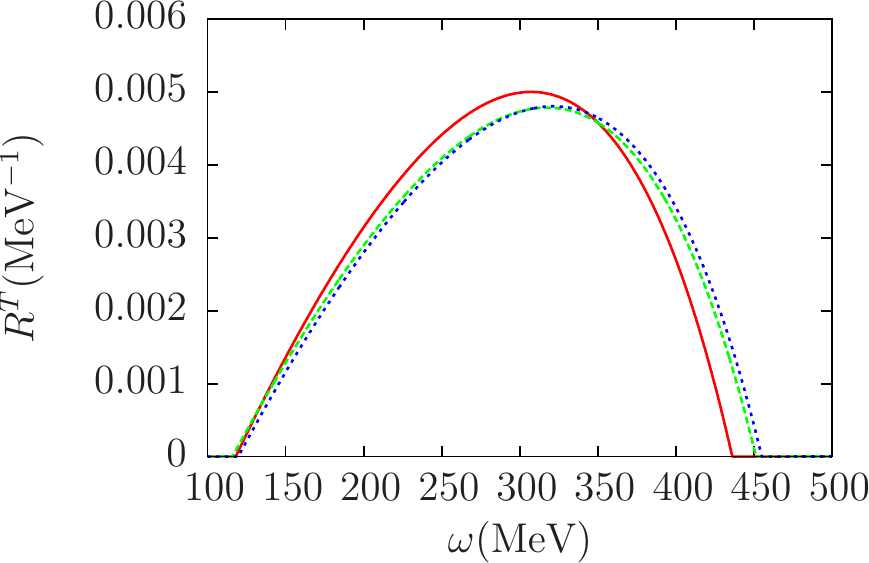}
\\
\includegraphics[scale=0.8, angle=0]{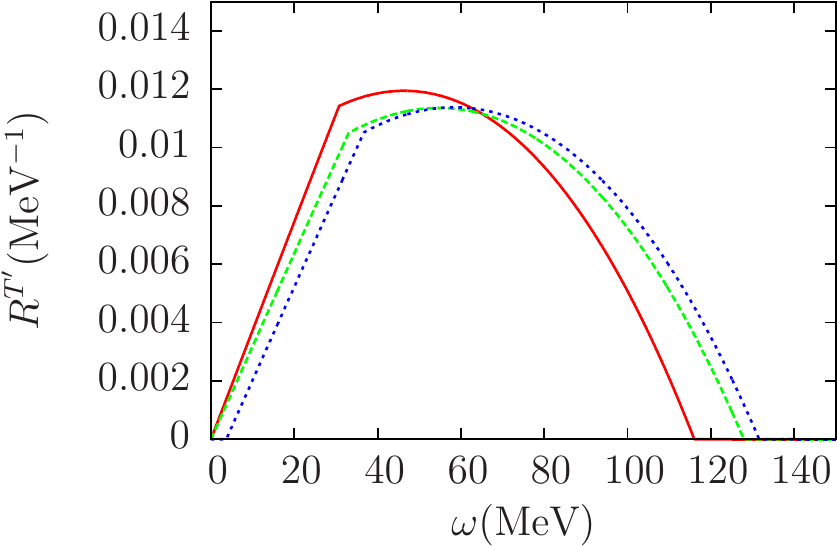}
\hspace{1.cm}
\includegraphics[scale=0.8, angle=0]{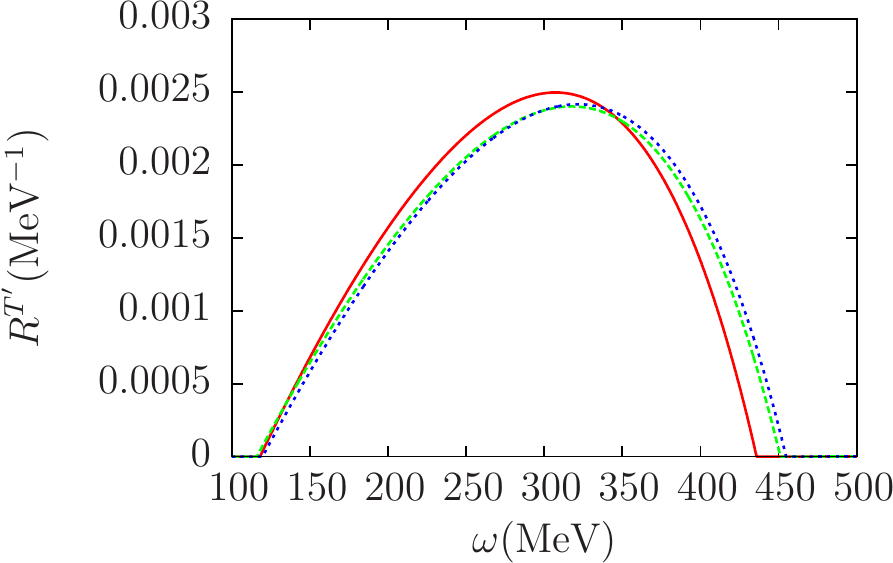}
\caption{(Color online) Neutrino charged-current weak response functions per neutron of $^{208}$Pb 
 in the symmetric (SRFG) and asymmetric (ARFG) relativistic Fermi gas. The ARFG results with no energy shift  (see text) are also shown.
 Each column corresponds to a fixed value of the momentum transfer $q$.}
\label{fig:Pbnumu} 
\end{figure}

\begin{figure}[h]
  \includegraphics[scale=0.8, angle=0]{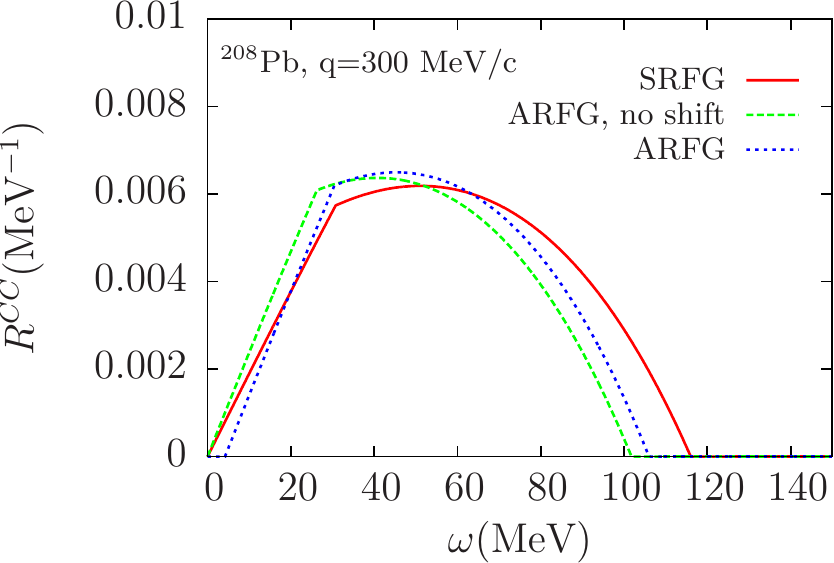}
\hspace{1.cm}
\includegraphics[scale=0.8, angle=0]{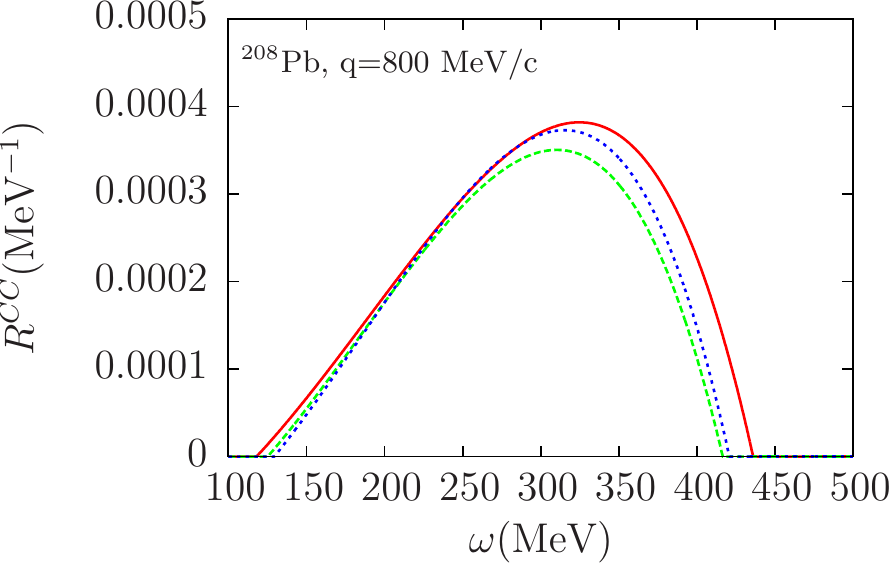}
\\
\includegraphics[scale=0.8, angle=0]{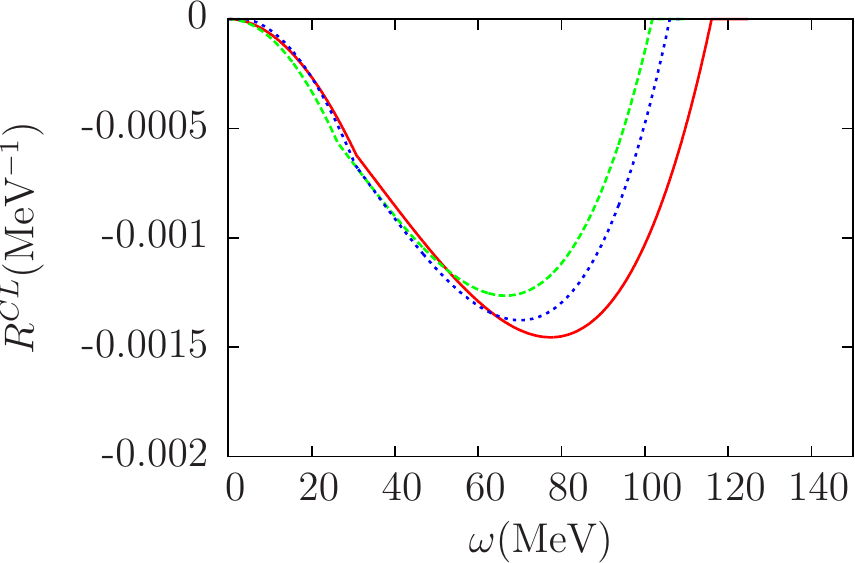}
\hspace{1.cm}
\includegraphics[scale=0.8, angle=0]{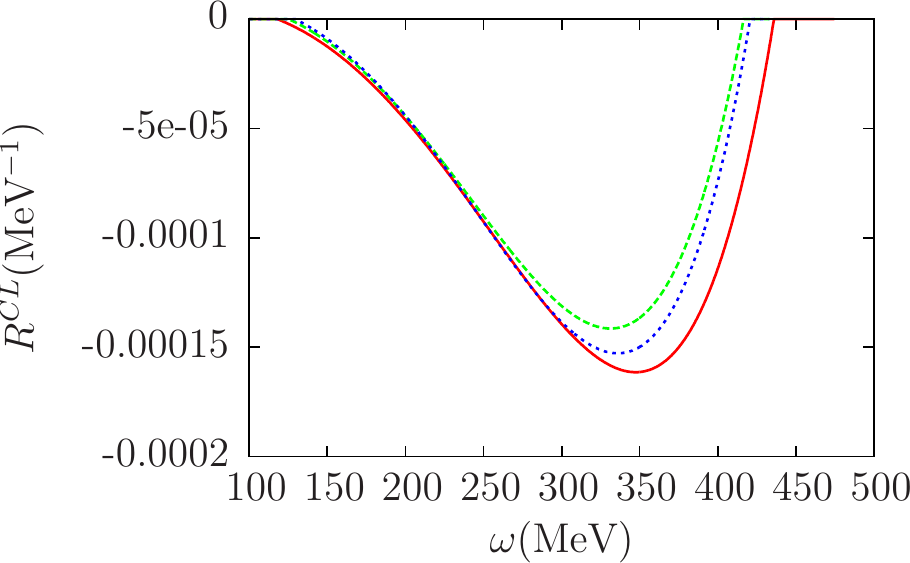}
\\
\includegraphics[scale=0.8, angle=0]{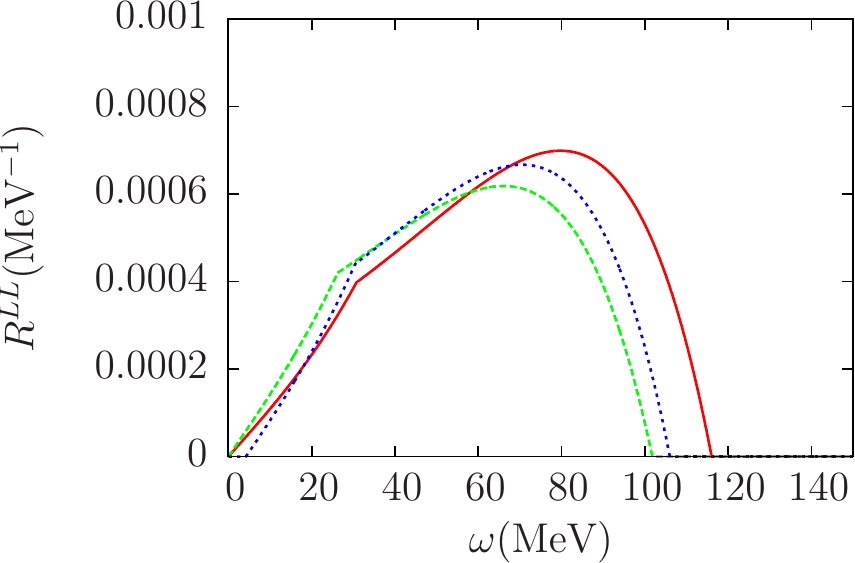}
\hspace{1.cm}
\includegraphics[scale=0.8, angle=0]{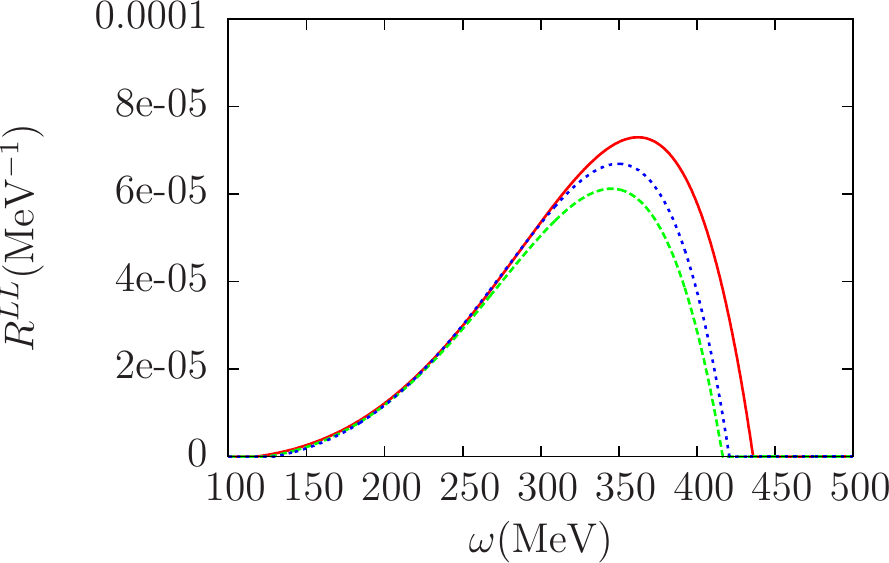}
\\
\includegraphics[scale=0.8, angle=0]{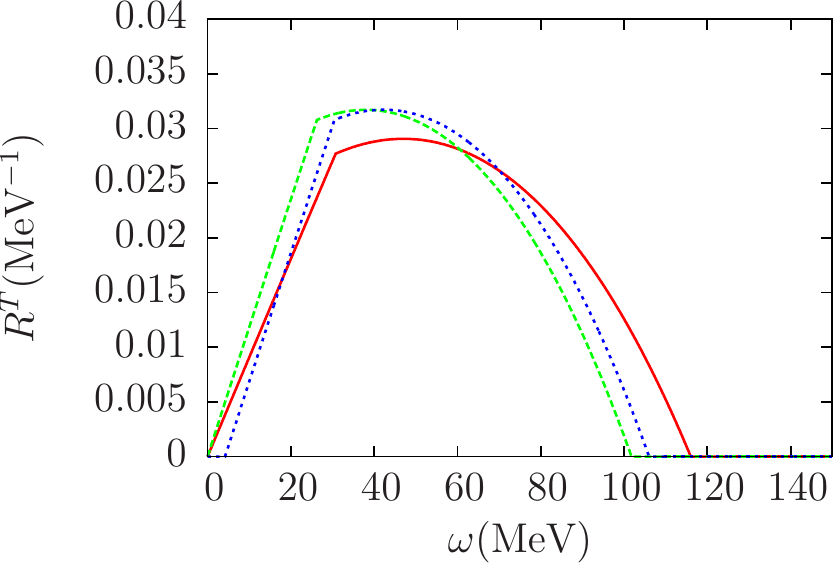}
\hspace{1.cm}
\includegraphics[scale=0.8, angle=0]{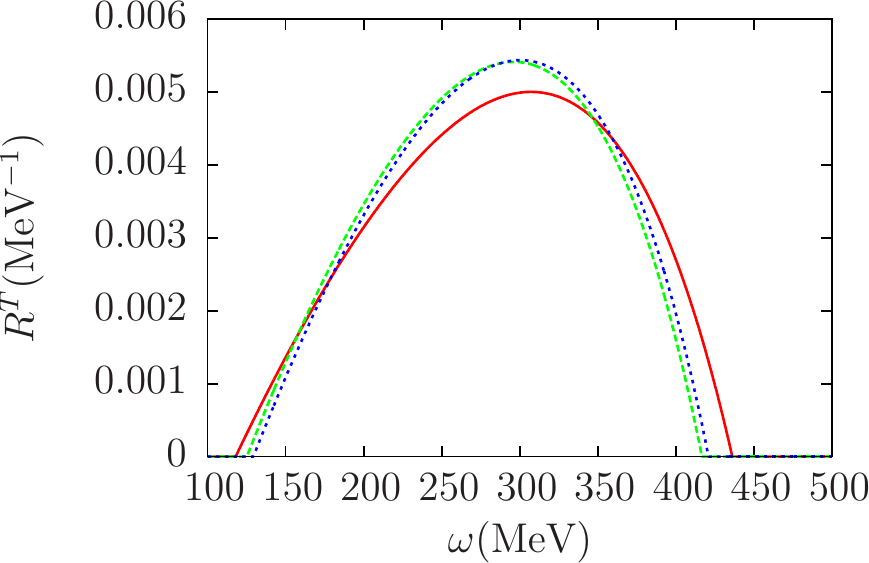}
\\
\includegraphics[scale=0.8, angle=0]{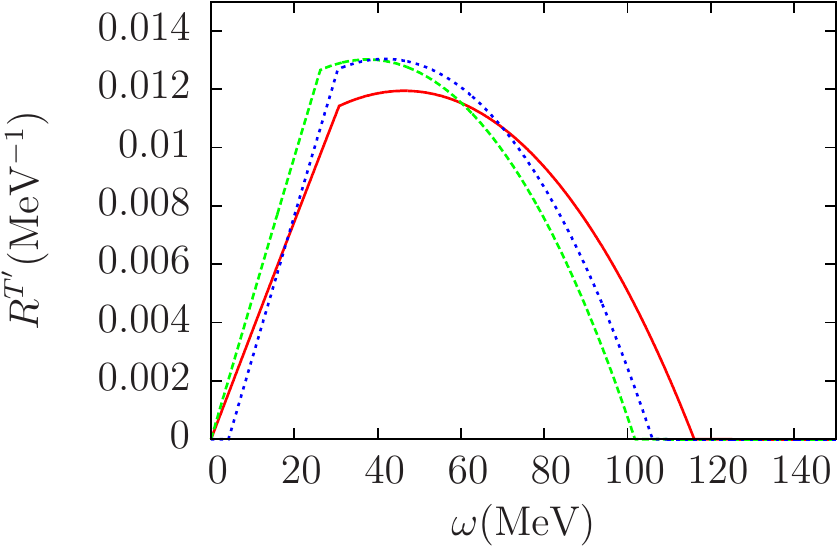}
\hspace{1.cm}
\includegraphics[scale=0.8, angle=0]{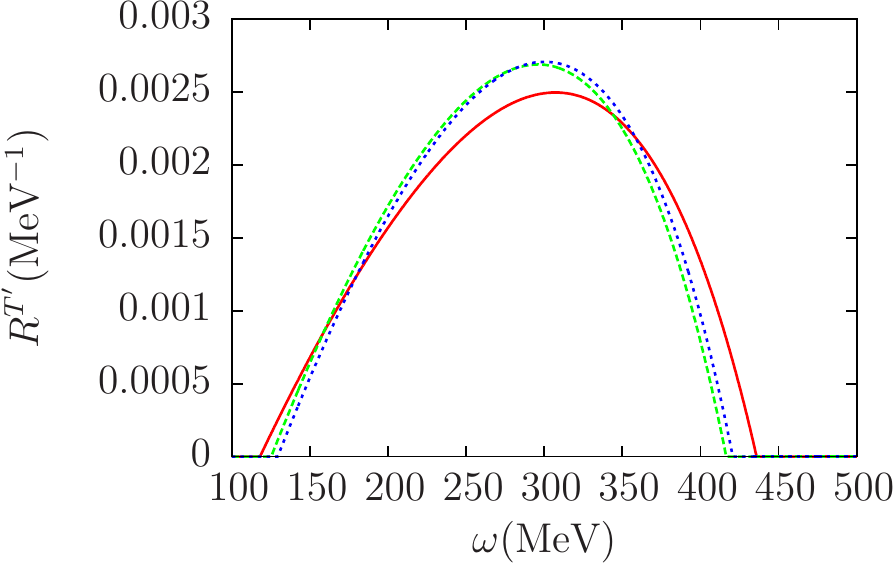}
\caption{(Color online) Antineutrino charged-current weak response functions per proton of $^{208}$Pb 
 in the symmetric (SRFG) and asymmetric (ARFG) relativistic Fermi gas. The ARFG results with no energy shift (see text) are also shown.
 Each column corresponds to a fixed value of the momentum transfer $q$.}
\label{fig:Pbnubarmu} 
\end{figure}

The numerical results for $(\nu_\mu,\mu^-)$ and $(\bar\nu_\mu,\mu^+)$ scattering corresponding to  carbon, argon and lead targets are shown in Figs. \ref{fig:Cnumu}-\ref{fig:Pbnubarmu}. Although not shown here, oxygen and iron, used as well in neutrino oscillation experiments, have also been considered. The results are very similar to those obtained for carbon and argon, respectively.

The main observation here is that the differences between the SRFG and ARFG results are much larger than was seen above for the NC processes. As noted in Sect.~\ref{subsec:NC}, for NC processes in the ARFG model the energy offsets cancel and the differences between the SRFG and ARFG arise entirely from the different Fermi momenta entering for protons and neutrons. In contrast, for charge-changing weak interactions this is not the case as the relative offsets for protons and neutrons (which are usually different) do enter. Hence the results corresponding to symmetric target nuclei, such as $^{12}$C and $^{16}$O, will also differ from the usual SRFG results, since the final nucleus will have $Z\ne N$.
In order to disentangle these two effects, we also show in each plot the results, labeled as ``ARFG, no shift", where the separation energies $S_p$ and $S_n$ are set to zero.

We observe that:

\begin{itemize}
\item for $^{12}$C 
the effect of the energy offsets is large, as could be anticipated looking at Table II, while the difference in $k_F$ plays a minor role; these effects are generally larger for neutrinos than for antineutrinos;
\item for $^{40}$Ar, which is relevant for neutrino oscillation studies, 
the two effects are comparable and both contribute to a slight shift of the responses to higher energy transfers, in particular for low $q$;
\item for $^{208}$Pb the main differences between SRFG and ARFG are due to Fermi momentum effects, which shifts the neutrino (antineutrino) responses to higher (lower) energy transfers.
\end{itemize}

\begin{table}[h]
  \begin{tabular}{| c | c | c | c | c | c |}
    \hline
   $(\nu_\mu,\mu^-)$      &  $CC$  & $CL$  &  $LL$  &  $T$  &  $T^\prime$  \\ \hline \hline
 $^{12}$C       &      1.24    &        1.31  & 1.37 &   1.02 & 1.03    \\ \hline
 $^{40}$Ar       &       1.06    &        1.08  &  1.11 & 0.98 & 0.98  \\ \hline
 $^{208}$Pb       &      1.15    &         1.21  & 1.27 & 0.96 & 0.97  \\ \hline 
  \end{tabular}
  \ \ \ \ \ \ \ 
  \begin{tabular}{| c | c | c | c | c | c |}
    \hline
   $(\bar\nu_\mu,\mu^+)$      &  $CC$  & $CL$  &  $LL$  &  $T$  &  $T^\prime$  \\ \hline \hline
 $^{12}$C       &      1.15    &        1.19  & 1.22 &   1.02 & 1.01    \\ \hline
 $^{40}$Ar       &       1.05    &        1.05  &  1.05 & 1.04 & 1.04  \\ \hline
 $^{208}$Pb       &      0.98    &         0.95  & 0.92 & 1.09 & 1.08  \\ \hline 
  \end{tabular}
\caption{Ratios between the ARFG and SRFG weak response functions at the QEP for $q$=800 MeV/c for neutrino (left table) and antineutrino (right table) CC scattering.
}
  \label{tab:ratios}
  \end{table}

As was done in the NC case, to get an idea of the importance of the asymmetry effects we list in Table \ref{tab:ratios} the ratios between the ARFG and SRFG responses  for neutrino and antineutrino scattering at $q$=800 MeV/c. We observe that the effects are minor for argon, while they are important for carbon and lead. Moreover, due to the different origin of the effects illustrated above, in the case of carbon the asymmetric model yields higher responses at the quasielastic peak for both neutrino and antineutrino scattering and in all five channels, while for lead the asymmetry effects have the opposite sign for neutrinos and antineutrinos and they depend on the channel: they increase (decrease) the neutrino (antineutrino) charge/longitudinal responses, whereas the opposite occurs for the transverse $T$ ant $T^\prime$ responses.

\section{Conclusions}
\label{sec:Concl}

In the present study we have developed an extension of the familiar relativistic Fermi gas model for inclusive semi-leptonic electroweak processes. Specifically, we have developed the model for neutral current scattering of electrons or (anti)neutrinos and for charge-changing (anti)neutrino reactions with nuclei. The new element in this work has been to allow for effects arising from differences between protons and neutrons within the context of the relativistic Fermi gas in order to evaluate how significant such effects may be in progressing from light $N=Z$ ({\it i.e.,} symmetric) nuclei to asymmetric nuclei having $N>Z$. We denote the usual relativistic Fermi gas model, typically abbreviated RFG, to be the symmetric relativistic Fermi gas (SRFG), while the new extension to asymmetric nuclei we denote as the asymmetric relativistic Fermi gas (ARFG).

Two types of extensions have been studied: first, we consider only nuclei in the valley of stability where typically the volume occupied by protons and neutrons in the ground states of such nuclei is the same, and hence where the densities scale by the numbers of neutrons and protons. In the context of the Fermi gas this implies that the Fermi momenta for $n$ and $p$ will be different, scaling by $(N/Z)^{1/3}$. Second, we adjust the Fermi energies of the proton and neutron gases of the parent nucleus, and its neighbors in the case of CC (anti)neutrino reactions, to agree with the measured values. We note that this is a basic assumption in the present ARFG model and not the only way one might proceed.  For instance, one might develop a different model where the energy offsets involved are allowed to be chosen by forcing agreement with experiment. The motivation in the present work is to explore the typical size of these second effects to see if they are typically negligible or if they should be taken seriously in future more sophisticated modeling. 

One conclusion is that the density effect (leading to different Fermi momenta for neutrons and protons) plays no role at all for NC scattering from symmetric nuclei and a relatively minor role for CC (anti)neutrino reactions in such systems where there is some effect since neighboring nuclei which have slightly different Fermi momenta are involved. In contrast, for very asymmetric nuclei such as $^{208}$Pb the effects from having differing neutron and proton Fermi momenta are somewhat larger, although still relatively minor, for NC scattering, but much more significant for CC (anti)neutrino reactions.

A second observation is that in NC scattering (electrons or neutrinos) the energy offsets do not play a role; simply put, only energy differences between particles and holes enter and, since the 1p1h states involve only excitations of protons or neutrons individually, the offsets cancel.  In contrast, for CC (anti)neutrino reactions neutrons change into protons or {\it vice versa} and thus the offsets do play a role. For the last type of reaction one sees that the energy offset effect is dominant in light symmetric nuclei such as $^{12}$C, roughly comparable to the density effect for $^{40}$Ar, and sub-dominant to the density effect in very asymmetric nuclei such as $^{208}$Pb.

The ARFG model developed in this study can be extended straightforwardly to include inelastic processes following previous work done along these lines for the SRFG. Finally, while much more involved than the 1p1h focus of the present work, it is possible to extend the previous 2p2h SRFG studies of two-body MEC contributions \cite{Simo:2016ikv,Amaro:2017eah} to incorporate asymmetric nuclei; such a study is in progress.

\begin{acknowledgments}

This work was partially supported by the INFN under project
MANYBODY, by the University of Turin under contract BARM-RILO-17, by
the Spanish Ministerio de Economia y Competitividad and ERDF (European Regional Development
Fund) under contracts FIS2014-59386-P, FIS2014-53448-C2-1, FIS2017-88410-P, by the Junta de
Andalucia (grants No. FQM-225, FQM160),  and part (TWD) by the U.S. Department of Energy under
cooperative agreement DE-FC02-94ER40818. MBB acknowledges support from the ``Emilie du Ch\^atelet" programme of the P2IO LabEx (ANR-10-LABX-0038).
GDM acknowledges support from a Junta de Andalucia fellowship (FQM7632, Proyectos de Excelencia 2011). JWVO acknowledges support by the US Department of Energy under Contract No. DE-AC05-06OR23177, and by the U.S. Department of Energy cooperative research agreement DE-AC05-84ER40150.
\end{acknowledgments}

\bibliography{biblio}

\begin{thebibliography}{14}
\expandafter\ifx\csname natexlab\endcsname\relax\def\natexlab#1{#1}\fi
\expandafter\ifx\csname bibnamefont\endcsname\relax
  \def\bibnamefont#1{#1}\fi
\expandafter\ifx\csname bibfnamefont\endcsname\relax
  \def\bibfnamefont#1{#1}\fi
\expandafter\ifx\csname citenamefont\endcsname\relax
  \def\citenamefont#1{#1}\fi
\expandafter\ifx\csname url\endcsname\relax
  \def\url#1{\texttt{#1}}\fi
\expandafter\ifx\csname urlprefix\endcsname\relax\def\urlprefix{URL }\fi
\providecommand{\bibinfo}[2]{#2}
\providecommand{\eprint}[2][]{\url{#2}}

\bibitem[{\citenamefont{Alberico et~al.}(1989)\citenamefont{Alberico, Drago,
  and Villavecchia}}]{Alberico:1989zz}
\bibinfo{author}{\bibfnamefont{W.~M.} \bibnamefont{Alberico}},
  \bibinfo{author}{\bibfnamefont{A.}~\bibnamefont{Drago}}, \bibnamefont{and}
  \bibinfo{author}{\bibfnamefont{C.}~\bibnamefont{Villavecchia}},
  \bibinfo{journal}{Nucl. Phys.} \textbf{\bibinfo{volume}{A505}},
  \bibinfo{pages}{309} (\bibinfo{year}{1989}).

\bibitem[{\citenamefont{Alberico et~al.}(1987)\citenamefont{Alberico, Chanfray,
  Ericson, and Molinari}}]{Alberico:1987zgx}
\bibinfo{author}{\bibfnamefont{W.~M.} \bibnamefont{Alberico}},
  \bibinfo{author}{\bibfnamefont{G.}~\bibnamefont{Chanfray}},
  \bibinfo{author}{\bibfnamefont{M.}~\bibnamefont{Ericson}}, \bibnamefont{and}
  \bibinfo{author}{\bibfnamefont{A.}~\bibnamefont{Molinari}},
  \bibinfo{journal}{Nucl. Phys.} \textbf{\bibinfo{volume}{A475}},
  \bibinfo{pages}{233} (\bibinfo{year}{1987}).

\bibitem[{\citenamefont{Davesne et~al.}(2014)\citenamefont{Davesne, Pastore,
  and Navarro}}]{Davesne:2014yaa}
\bibinfo{author}{\bibfnamefont{D.}~\bibnamefont{Davesne}},
  \bibinfo{author}{\bibfnamefont{A.}~\bibnamefont{Pastore}}, \bibnamefont{and}
  \bibinfo{author}{\bibfnamefont{J.}~\bibnamefont{Navarro}},
  \bibinfo{journal}{Phys. Rev.} \textbf{\bibinfo{volume}{C89}},
  \bibinfo{pages}{044302} (\bibinfo{year}{2014}).

\bibitem[{\citenamefont{Lipparini and Pederiva}(2013)}]{Lipparini:2013fma}
\bibinfo{author}{\bibfnamefont{E.}~\bibnamefont{Lipparini}} \bibnamefont{and}
  \bibinfo{author}{\bibfnamefont{F.}~\bibnamefont{Pederiva}},
  \bibinfo{journal}{Phys. Rev.} \textbf{\bibinfo{volume}{C88}},
  \bibinfo{pages}{024318} (\bibinfo{year}{2013}).

\bibitem[{\citenamefont{Lipparini and Pederiva}(2016)}]{Lipparini:2016fyb}
\bibinfo{author}{\bibfnamefont{E.}~\bibnamefont{Lipparini}} \bibnamefont{and}
  \bibinfo{author}{\bibfnamefont{F.}~\bibnamefont{Pederiva}},
  \bibinfo{journal}{Phys. Rev.} \textbf{\bibinfo{volume}{C94}},
  \bibinfo{pages}{024323} (\bibinfo{year}{2016}).

\bibitem[{\citenamefont{Alvarez-Ruso et~al.}(2018)}]{Alvarez-Ruso:2017oui}
\bibinfo{author}{\bibfnamefont{L.}~\bibnamefont{Alvarez-Ruso}}
  \bibnamefont{et~al.}, \bibinfo{journal}{Prog. Part. Nucl. Phys.}
  \textbf{\bibinfo{volume}{100}}, \bibinfo{pages}{1} (\bibinfo{year}{2018}).

\bibitem[{\citenamefont{Barbaro et~al.}(2004)\citenamefont{Barbaro, Caballero,
  Donnelly, and Maieron}}]{PhysRevC.69.035502}
\bibinfo{author}{\bibfnamefont{M.~B.} \bibnamefont{Barbaro}},
  \bibinfo{author}{\bibfnamefont{J.~A.} \bibnamefont{Caballero}},
  \bibinfo{author}{\bibfnamefont{T.~W.} \bibnamefont{Donnelly}},
  \bibnamefont{and} \bibinfo{author}{\bibfnamefont{C.}~\bibnamefont{Maieron}},
  \bibinfo{journal}{Phys. Rev. C} \textbf{\bibinfo{volume}{69}},
  \bibinfo{pages}{035502} (\bibinfo{year}{2004}).

\bibitem[{\citenamefont{Cenni et~al.}(1977)\citenamefont{Cenni, Donnelly, and
  Molinari}}]{Cenni:1996zh}
\bibinfo{author}{\bibfnamefont{R.}~\bibnamefont{Cenni}},
  \bibinfo{author}{\bibfnamefont{T.~W.} \bibnamefont{Donnelly}},
  \bibnamefont{and} \bibinfo{author}{\bibfnamefont{A.}~\bibnamefont{Molinari}},
  \bibinfo{journal}{Phys. Rev.} \textbf{\bibinfo{volume}{C56}},
  \bibinfo{pages}{276} (\bibinfo{year}{1977}).

\bibitem[{\citenamefont{Maieron et~al.}(2002)\citenamefont{Maieron, Donnelly,
  and Sick}}]{PhysRevC.65.025502}
\bibinfo{author}{\bibfnamefont{C.}~\bibnamefont{Maieron}},
  \bibinfo{author}{\bibfnamefont{T.~W.} \bibnamefont{Donnelly}},
  \bibnamefont{and} \bibinfo{author}{\bibfnamefont{I.}~\bibnamefont{Sick}},
  \bibinfo{journal}{Phys. Rev. C} \textbf{\bibinfo{volume}{65}},
  \bibinfo{pages}{025502} (\bibinfo{year}{2002}).

\bibitem[{\citenamefont{Tepel}(1984)}]{TEPEL1984129}
\bibinfo{author}{\bibfnamefont{J.}~\bibnamefont{Tepel}},
  \bibinfo{journal}{Computer Physics Communications}
  \textbf{\bibinfo{volume}{33}}, \bibinfo{pages}{129 } (\bibinfo{year}{1984}),
  ISSN \bibinfo{issn}{0010-4655}.

\bibitem[{\citenamefont{Amaro et~al.}(2005)\citenamefont{Amaro, Barbaro,
  Caballero, Donnelly, Molinari, and Sick}}]{Amaro:2004bs}
\bibinfo{author}{\bibfnamefont{J.~E.} \bibnamefont{Amaro}},
  \bibinfo{author}{\bibfnamefont{M.~B.} \bibnamefont{Barbaro}},
  \bibinfo{author}{\bibfnamefont{J.~A.} \bibnamefont{Caballero}},
  \bibinfo{author}{\bibfnamefont{T.~W.} \bibnamefont{Donnelly}},
  \bibinfo{author}{\bibfnamefont{A.}~\bibnamefont{Molinari}}, \bibnamefont{and}
  \bibinfo{author}{\bibfnamefont{I.}~\bibnamefont{Sick}},
  \bibinfo{journal}{Phys. Rev. C} \textbf{\bibinfo{volume}{71}},
  \bibinfo{pages}{015501} (\bibinfo{year}{2005}).

\bibitem[{\citenamefont{Amaro et~al.}(2006)\citenamefont{Amaro, Barbaro,
  Caballero, and Donnelly}}]{PhysRevC.73.035503}
\bibinfo{author}{\bibfnamefont{J.~E.} \bibnamefont{Amaro}},
  \bibinfo{author}{\bibfnamefont{M.~B.} \bibnamefont{Barbaro}},
  \bibinfo{author}{\bibfnamefont{J.~A.} \bibnamefont{Caballero}},
  \bibnamefont{and} \bibinfo{author}{\bibfnamefont{T.~W.}
  \bibnamefont{Donnelly}}, \bibinfo{journal}{Phys. Rev. C}
  \textbf{\bibinfo{volume}{73}}, \bibinfo{pages}{035503}
  (\bibinfo{year}{2006}).

\bibitem[{\citenamefont{Ruiz~Simo et~al.}(2017)\citenamefont{Ruiz~Simo, Amaro,
  Barbaro, De~Pace, Caballero, and Donnelly}}]{Simo:2016ikv}
\bibinfo{author}{\bibfnamefont{I.}~\bibnamefont{Ruiz~Simo}},
  \bibinfo{author}{\bibfnamefont{J.~E.} \bibnamefont{Amaro}},
  \bibinfo{author}{\bibfnamefont{M.~B.} \bibnamefont{Barbaro}},
  \bibinfo{author}{\bibfnamefont{A.}~\bibnamefont{De~Pace}},
  \bibinfo{author}{\bibfnamefont{J.~A.} \bibnamefont{Caballero}},
  \bibnamefont{and} \bibinfo{author}{\bibfnamefont{T.~W.}
  \bibnamefont{Donnelly}}, \bibinfo{journal}{J. Phys.}
  \textbf{\bibinfo{volume}{G44}}, \bibinfo{pages}{065105}
  (\bibinfo{year}{2017}).

\bibitem[{\citenamefont{Amaro et~al.}(2017)\citenamefont{Amaro, Barbaro,
  Caballero, De~Pace, Donnelly, Megias, and Ruiz~Simo}}]{Amaro:2017eah}
\bibinfo{author}{\bibfnamefont{J.~E.} \bibnamefont{Amaro}},
  \bibinfo{author}{\bibfnamefont{M.~B.} \bibnamefont{Barbaro}},
  \bibinfo{author}{\bibfnamefont{J.~A.} \bibnamefont{Caballero}},
  \bibinfo{author}{\bibfnamefont{A.}~\bibnamefont{De~Pace}},
  \bibinfo{author}{\bibfnamefont{T.~W.} \bibnamefont{Donnelly}},
  \bibinfo{author}{\bibfnamefont{G.~D.} \bibnamefont{Megias}},
  \bibnamefont{and}
  \bibinfo{author}{\bibfnamefont{I.}~\bibnamefont{Ruiz~Simo}},
  \bibinfo{journal}{Phys. Rev.} \textbf{\bibinfo{volume}{C95}},
  \bibinfo{pages}{065502} (\bibinfo{year}{2017}).

\end{thebibliography}

\end{document}